\documentclass[12pt,preprint]{aastex}

\usepackage{graphicx}

\def\Mdot{\hbox{$\dot {M}$}}

\def\Rsun{\hbox{\it R$_\odot$}}
\def\Zsun{\hbox{\it Z$_\odot$}}
\def\Rstar{\hbox{\it R$_*$}}
\def\Lsun{\hbox{\it L$_\odot$}}
\def\Lstar{\hbox{\it L$_*$}}
\def\Msun{\hbox{\it M$_\odot$}}
\def\Minit{\hbox{\it M$_{\rm initial}$}}
\def\Msunyr{\hbox{\it M$_\odot\,$yr$^{-1}$}}
\def\Myr{\hbox{\it Myr}}

\def\Teff{\hbox{\it T$_{\rm eff}$}}
\def\Vinf{\hbox{$v_\infty$}}
\def\kms{\hbox{km$\,$s$^{-1}$}}

\def\AH{\hbox{\it A$_{\rm H}$}}
\def\AK{\hbox{\it A$_{\rm K}$}}
\def\H{\hbox{\it H}}
\def\K{\hbox{\it K}}

\def\simgr{\mathrel{\hbox{\rlap{\hbox{\lower4pt\hbox{$\sim$}}}\hbox{$>$}}}}
\def\mnk{\hbox{\it m$_{\rm F205W}$}}
\def\mnh{\hbox{\it m$_{\rm F160W}$}}

\def\Fnp{\hbox{\it F$_{\rm F187N}$}}
\def\Fnc{\hbox{\it F$_{\rm F190N}$}}
\def\Frh{\hbox{\it F$_{\rm 8.5~GHz}$}}
\def\Frl{\hbox{\it F$_{\rm 4.9~GHz}$}}
\def\ew{\hbox{\it EW$_{\rm 1.87~\micron}$}}

\makeatletter
\def\jnl@aj{ApJ}
\ifx\revtex@jnl\jnl@aj\let\tablebreak=\\\fi
\makeatother

\received{}
\revised{REVISION DATE}
\accepted{}
\journalid{VOL}{JOURNAL DATE}
\articleid{START PAGE}{END PAGE}
\paperid{MANUSCRIPT ID}
\cpright{TYPE}{YEAR}
\ccc{CODE}

\shorttitle{Arches Cluster}
\shortauthors{Figer et al.}

\begin{document}

\title{Massive Stars in the Arches Cluster
\footnote{Based on observations with the 
NASA/ESA Hubble Space Telescope, obtained
at the Space Telescope Science Institute, which is operated by the Association of Universities
for Research in Astronomy, Inc. under NASA contract No. NAS5-26555.}
\footnote{Data presented herein were obtained at the W.M. Keck
Observatory,  which is operated as a scientific partnership among the
California Institute of Technology,  the University of California and
the National Aeronautics and Space  Administration.  The Observatory
was made possible by the generous financial support of the W.M. Keck
Foundation.}}

\author{
Donald F. Figer\altaffilmark{3,4}, Francisco Najarro\altaffilmark{5}, Diane Gilmore\altaffilmark{3}, \\
Mark Morris\altaffilmark{6}, Sungsoo S. Kim\altaffilmark{6}, Eugene Serabyn\altaffilmark{7}, \\
Ian S. McLean\altaffilmark{6}, Andrea M. Gilbert\altaffilmark{8}, James R. Graham\altaffilmark{8}, \\
James E. Larkin\altaffilmark{6}, N. A. Levenson\altaffilmark{4}, Harry I. Teplitz\altaffilmark{9,}\altaffilmark{10}}

\email{figer@stsci.edu}

\altaffiltext{3}{Space Telescope Science Institute, 3700 San Martin Drive, Baltimore, MD 21218; figer@stsci.edu }
\altaffiltext{4}{Department of Physics and Astronomy, Johns Hopkins University, Baltimore, MD  21218}
\altaffiltext{5}{Instituto de Estructura de la Materia, CSIC, Serrano 121, 29006 Madrid, Spain }
\altaffiltext{6}{Department of Physics and Astronomy, University of California, Los Angeles, 
	Division of Astronomy, Los Angeles, CA, 90095-1562 }
\altaffiltext{7}{Caltech, 320-47, Pasadena, CA 91125; eserabyn@huey.jpl.nasa.gov}
\altaffiltext{8}{Department of Astronomy, University of California, Berkeley, 601 Campbell Hall, Berkeley, CA, 94720-3411}
\altaffiltext{9}{Laboratory for Astronomy and Solar Physics, Code 681, Goddard Space Flight Center, Greenbelt MD 20771}
\altaffiltext{10}{NOAO Research Associate}

\begin{abstract}
We present and use new spectra and narrow-band images, 
along with previously published broad-band images, of stars in the Arches cluster to extract photometry, astrometry, equivalent
width, and velocity information. The data are interpreted with a wind/atmosphere code
to determine stellar temperatures, luminosities, mass-loss rates, and abundances. 
We have doubled the number of known emission-line stars, and 
we have also made the first spectroscopic identification of the main sequence for any population in 
the Galactic Center.
We conclude that the most massive stars are bona-fide Wolf-Rayet (WR) stars and are some of the most massive stars known,
having \Minit~$>$100~\Msun, and prodigious winds, \Mdot~$>$10$^{-5}$~\Msunyr, that are 
enriched with helium and nitrogen; with these identifications, the Arches cluster
contains about 5\% of all known WR stars in the Galaxy. 
We find an upper limit to the velocity dispersion of 22~\kms, implying an
upper limit to the cluster mass of 7(10$^4$)~\Msun\ within a radius of 0.23~pc; we also 
estimate the bulk heliocentric velocity of the cluster
to be v$_{\rm cluster,\odot}\approx+95$~\kms. 
Taken together, these results suggest that the Arches cluster was formed in a short, but massive, burst of
star formation about 2.5$\pm$0.5~\Myr\ ago, from a molecular
cloud which is no longer present. The cluster happens to be approaching and ionizing the surface of a background 
molecular cloud, thus producing the Thermal Arched Filaments. We estimate that the cluster produces
4(10$^{51}$)~ionizing photons~s$^{-1}$, more than enough to account for the observed 
thermal radio flux from the nearby cloud, 3(10$^{49}$)~ionizing photons~s$^{-1}$.
Commensurately, it produces 10$^{7.8}$~\Lsun\ in total luminosity, providing the
heating source for the nearby molecular cloud, L$_{\rm cloud}\approx10^7$~\Lsun. 
These interactions between a cluster of hot
stars and a wayward molecular cloud are similar to those seen in the ``Quintuplet/Sickle''
region. The small spread of formation times for the known young
clusters in the Galactic Center, and the relative lack of intermediate-age stars 
($\tau_{\rm age}$=10$^{7.0}$ to 10$^{7.3}$~yrs),
suggest that the Galactic Center has recently been host to a burst of star formation.
Finally, we have made new identifications of near-infrared sources that are counterparts
to recently identified x-ray and radio sources.
\end{abstract}


\section{Introduction}
The Arches cluster is an extraordinarily massive and dense young cluster of stars near
the Galactic Center. First discovered about 10 years ago as a compact collection of a 
dozen or so emission-line stars \citep{cot92,nag95,fig95a,cot95,cot96}, 
the cluster contains thousands of stars, including at least 160 O stars  \citep{ser98,fig99a}. 
 \citet{fig99a} used HST/NICMOS observations to estimate a total cluster 
mass ($\gtrsim$10$^4$~\Msun) and radius (0.2~pc) to arrive at an average mass density of 
3(10$^5$)~\Msun~pc$^{-3}$ in stars, suggesting that the Arches cluster is the densest, 
and one of the most 
massive, young clusters in the Galaxy. They further used these data to estimate an initial 
mass function (IMF) which is very flat ($\Gamma$~$\sim-$0.6$\pm$0.1) with respect to what has been 
found for the solar neighborhood \citep[$\Gamma$~$\sim-$1.35]{sal55}
and other Galactic clusters \citep{sca98}. They also estimated
an age of 2$\pm$1~\Myr, based on the magnitudes, colors, and mix of spectral types, which
makes the cluster ideal for testing massive stellar-evolution models.   

Given its extraordinary nature, the Arches cluster has been a target for many new
observations. \citet{sto02} recently verified a flat IMF slope for the Arches cluster,
finding $\Gamma = -0.8$, using both adaptive optics imaging with the Gemini North telescope and 
the HST/NICMOS data presented in \citet{fig99a}. 
\citet{blu01} used adaptive optics imaging at the CFHT and HST/NICMOS data 
(also presented in this paper) to identify several 
new emission-line stars and estimate an age for the cluster of 2$-$4.5~\Myr.
\citet{lan01a} detected eight radio sources, seven of which have thermal spectral indices and stellar counterparts, within 10\arcsec\
of the center of the cluster. They suggest that the stellar winds from the counterparts produce the radio emission
via free-free emission, consistent with earlier indications from near-infrared
narrow-band imaging \citep{nag95} and spectroscopy \citep{cot96}. In a related study, 
\citet{lan01b} argued that the hot stars in the Arches cluster are responsible for ionizing
the surface of a nearby molecular cloud to produce the arches filaments, as originally
suggested by \citet{cot96} and \citet{ser98}, but in contrast to earlier suggestions \citep{mor89,dav94,col96}.
\citet{zad01} used the Chandra telescope to detect
three x-ray components that they associate with the cluster, claiming that hot (10$^7$~K)
x-ray emitting gas is produced by an interaction between material expelled by the massive
stellar winds and the local interstellar medium. 

The Arches cluster has also been the target of several theoretical
studies regarding dynamical evolution of compact young clusters.  \citet{kim99} used
Fokker-Planck  models and  \citet{kim00} used N-body models to simulate 
the Arches cluster, assuming the presence of the
gravitational field of the Galactic Center. They found that such a cluster will disperse
through two-body interactions over a 10~\Myr\ timescale. 
 \citet{zwa01a} performed a similar study and found a similar result, although
they note the possibility that the Arches cluster is located in front of the plane containing
the Galactic Center. Finally, \citet{ger01} considered the possibility that
compact clusters formed outside the central parsec will plunge into the Galactic center as a result of
dynamical friction, eventually becoming similar in appearance to the young cluster currently residing there;
\citet{kim02} further consider this possibility.

In this paper, we use new and existing observations to determine the stellar properties of
the most massive stars in the Arches cluster. 
We present astrometry and photometry of stars with estimated initial masses
greater than 20~\Msun\ (the theoretical minimum mass of O stars), based upon HST/NICMOS
narrow-band and broad-band imaging.
We also present {\it K}-band high-resolution spectra of the emission-line stars, based upon Keck/NIRSPEC 
observations. We couple these data with previously-reported radio and x-ray data to infer 
stellar wind/atmosphere properties using a modeling code. Finally, we compare our results to
those reported in recent observational and theoretical papers. 

\section{Observations}
A log of our observations obtained using HST and KECK is given in Table~1.

\subsection{HST}
The HST data were obtained as part of GO-7364 (PI Figer), a program designed to measure the IMF's in the Arches
and Quintuplet clusters \citep{fig99a}, determine the star-formation history of the Galactic
Center \citep{ser02}, and determine the nature of the ``Pistol Star'' \citep{fig99b}.

Broad-band images were obtained using {\it HST}/NICMOS on UT 1997 September 13/14, 
in a 2$\times$2 mosaic pattern in the NIC2 aperture (19\farcs2 on a side).
Four nearby fields, separated from the center of the mosaic by 59\arcsec\ in a symmetric cross-pattern, 
were imaged in order to sample the background population. All fields were imaged in 
F110W ($\lambda_{\rm center}$~=~1.10~\micron), F160W ($\lambda_{\rm center}$~=~1.60~\micron), and 
F205W ($\lambda_{\rm center}$~=~2.05~\micron). The STEP256 sequence was used in 
the MULTIACCUM read mode with 11 reads, giving an exposure time of $\approx$256 seconds per image.
The plate scale was 0\farcs076 pixel$^{-1}$ (x) by 0\farcs075 pixel$^{-1}$ (y), in detector coordinates. 
The mosaic was centered on 
RA 17$^{\rm h}$45$^{\rm m}$50$\fs$35, DEC $-$28$^{\arcdeg}$49$^{\arcmin}$21$\farcs$82 (J2000), and 
the pattern orientation was $-$134\fdg6. The spectacular F205W image is shown in Figure~1a, after
first being processed with the standard STScI pipeline procedures.

The narrow-band images were obtained at roughly the same time as the broad-band images. One image
was obtained in each of the F187N ($\lambda_{\rm center}$~=~1.87~\micron) 
and F190N ($\lambda_{\rm center}$~=~1.90~\micron) filters in the NIC2 aperture. The filters widths are 
0.0194~\micron\ for F187N and 0.0177~\micron\ for F190N.
We used the same exposure parameters for these images as those used for the broad-band images.
The difference image, F187N minus F190N, is shown in Figure~2.

\subsection{Keck}
The spectroscopic observations were obtained on July 4, 1999, using NIRSPEC, the
facility near-infrared spectrometer, on the Keck II
telescope  \citep{mcl98,mcl02}, in high
resolution mode, covering {\it K}-band wavelengths ($1.98~\micron$ to
$2.28~\micron$). The long slit (24\arcsec) was positioned in a north-south
orientation on the sky, and a slit scan covering a 24\arcsec$\times$14\arcsec\ rectangular region 
was made by offsetting the telescope by a fraction of a slit width to the west between
successive exposures. The slit-viewing
camera (SCAM) was used to obtain images simultaneously with the
spectroscopic exposures, making it easy to determine the slit orientation
on the sky when the spectra were obtained. From SCAM images, we
estimate seeing (FWHM) of 0$\farcs$6. The plate scales for both spectrometer and SCAM
were taken from  \citet{fig00a}. We
chose to use the 3-pixel-wide slit (0\farcs43) in order to
match the FWHM of the seeing disk. The corresponding resolving
power was R$\sim$23,300 (=$\lambda/\Delta\lambda_{\rm FWHM}$), as measured from unresolved arc lamp lines. 

The NIRSPEC cross-disperser and the NIRSPEC-6 filter were used to
image six echelle orders onto the 1024$^2$-pixel InSb detector. 
The approximate spectral range covered in these orders
is listed in Table~2. Coverage includes \ion{He}{1} (2.058$~\micron$), \ion{He}{1} 
(2.112/113~\micron), Br$\gamma$/\ion{He}{1} (2.166$~\micron$),
\ion{He}{2} (2.189$~\micron$), \ion{N}{3} (2.24/25$~\micron$), and the CO bandhead, starting at
2.294$~\micron$ and extending to longer wavelengths beyond the range of the observations.

Quintuplet Star \#3  (hereafter ``Q3''), which is featureless in this
spectral region  \citep[Figure~1]{fig98},
was observed as a telluric standard  \citep{mon94}. Arc lamps containing Ar, Ne, Kr, and
Xe, were observed to set the wavelength scale. In addition, a continuum
lamp was observed through a vacuum gap etalon filter in order to produce an
accurate wavelength scale between arc lamp lines and sky lines
(predominantly from OH). A field relatively devoid of stars 
(RA~17$^{\rm h}$~44$^{\rm m}$~49$\fs$8,
DEC~$-$28$^{\arcdeg}$~54$^{\arcmin}$~6$\farcs$8~, J2000) was observed
to provide a dark current plus bias plus background image; this image was subtracted
from each target image. A quartz-tungsten-halogen (QTH) lamp was observed to provide a ``flat''
image which was divided into the background-subtracted target images. 

\section{Data Reduction}

\subsection{Photometry}
The NICMOS data were reduced as described in  \citet{fig99a}
using STScI pipeline routines, calnica and calnicb, and the most up-to-date
reference files. Star-finding, PSF-building, and PSF-fitting procedures were performed
using the DAOPHOT package  \citep{ste87} within the Image Reduction and 
Analysis Facility (IRAF)\footnote[10]
{IRAF is distributed by the National Optical Astronomy Observatories,
which are operated by the Association of Universities for Research
in Astronomy, Inc., under cooperative agreement with the National
Science Foundation.}. For the narrow-band photometry, 
PSF standard stars were identified in the field and
used in ALLSTAR. We used these stars to construct a model PSF with a radius of 15 pixels ($1.125\arcsec$).
This model was then fitted stars found throughout the image using DAOFIND.
Aperture corrections were estimated by comparing the magnitudes of the
PSF stars with those from an aperture of radius 7.5 pixels ($0.563\arcsec$),
and then adding $-2.5 \log (1.159)$ in order to extrapolate to an infinite aperture (M. Rieke, priv. comm.).
Table~1 gives the net aperture corrections to correct the aperture from a 3 pixel radius to infinity.

\subsection{Source Identification and Astrometry}
We culled the list produced by the process above by excluding stars with \AK$<$2.8 or \AK$>$4.2, or
equivalently, stars with \mnh$-$\mnk$<$1.4 or \mnh$-$\mnk$>$2.1. These choices are motivated
by the fact that the majority of stars in the Arches cluster have values within these
limits, as can be seen in Figure~4 of \citet{fig99a}; stars with values outside of these
limits are likely to be foreground or background stars.
The resultant star list is shown in 
Table~3. Objects are sorted according to inferred absolute {\it K}-band magnitudes, in order of decreasing brightness.
{\it K}-band absolute magnitudes were calculated using \AK\ = E$_{\rm H-K}/(\AH/\AK-1)$, 
where E$_{\rm H-K}=(H-K)-(H-K)_0$, and A$_{\lambda}\propto\lambda^{-1.53}$ \citep{rie89}.
We estimated intrinsic colors by convolving filter profiles with our best-matched model
spectra. In cases where spectra were not available, i.e.\ for faint stars, \AK~=~3.1, 
a value that is supported by \citet{fig99a}.
This absolute magnitude was then translated into an initial mass according to the procedure in \citet{fig99a}, except
that we assumed solar metallicity, a decision supported by our quantitative spectroscopic
analysis described later. 
Alternate identifications were taken from the following:
 \citet{nag95}, \citet{cot96}, \citet{lan01a}, and \citet{blu01}. We had to allow as much as a
1\arcsec\ offset in some cases so that the correct objects coincided in the various data sets. The position offsets
in the table are in right ascension (RA) and declination (Dec) and with respect to the star with the brightest inferred absolute
magnitude at K. These offsets were calculated by applying
the anamorphic plate scale at the time of observation and rotating x- and y- pixel
offsets in the focal plane in to RA and Dec offsets in the sky. 
The stars are plotted and numbered in Figure~1b.
Note that the masses in this table apply only in the case that
the stars satisfy our model assumptions. This is not the case for some of the stars, especially
the faint ones, given that they are likely to be field stars in the Galactic Center, but otherwise
unassociated with the Arches cluster. 

\subsection{Narrow-band Imaging}
The narrow-band filters cover wavelength regions that include 
several potentially relevant atomic transitions. 
There are potential contributions to the total observed flux through the F187N and F190N
filters from 8 and 2 transitions (\ion{H}{1} Paschen$-\alpha$, \ion{He}{1}, and \ion{He}{2}), respectively.
It is clear from the broad feature near 2.166~\micron\ that \ion{He}{1} lines are
strong in the spectra of the emission-line stars, while the relatively weak 2.189~\micron\
line indicates that the \ion{He}{2} lines falling in the F187N filter are minor
contributors to \Fnp. We estimate that the flux in the two lines falling in the F190N filter
is negligible. 

\subsection{Narrow-band equivalent-widths}
The equivalent-widths in Table~3 were computed according to the following equation:
\begin{equation}
EW_{1.87~\micron}~=~\frac{(F_{\rm F187N} - F_{\rm F190N})}{F_{\rm F190N}} \times \Delta \lambda,
\end{equation}
where the fluxes are in W\,cm$^{-2}$\,\micron$^{-1}$ and $\Delta\lambda$ is the 
FWHM of the F187N filter; this equation assumes that 
the emission line(s) lie completely within the filter bandwidth, and that there is no contamination
of \Fnc\ by emission or absorption lines. We increased \Fnp\
to account for the difference in reddening for the two wavelengths, assuming the extinction law of \citet{rie89};
this correction is approximately 8.0\% and depends slightly on the estimated extinction. 
In addition, we reduced \Fnp\ by 9.4\%
to account for the fact that the F187N filter has a shorter wavelength than the
F190N filter and thus \Fnp\ will be greater by this amount than \Fnc\ for normal
stars due to increasing flux toward shorter wavelengths on the Rayleigh-Jeans tail of the flux
distributions at these wavelengths. Apparently, these two effects nearly cancel each other. 

The difference image (F187N$-$F190N) in Figure~2 contains about two dozen emission-line
stars with significant flux excesses in the F187N filter, as listed in Table~3. 
The narrow-band equivalent-widths increase with apparent brightness, 
as seen in Figure~3. In addition, the colors are redder as a function of increasing
brightness \citep[Figure~4]{fig99a}. Both effects are consistent with the notion that the winds are radiatively
driven, so that higher luminosity stars will have stronger winds, and thus stronger
emission lines, with commensurately stronger free-free emission which has a relatively
flat (red) spectrum. In addition, the emission lines themselves
contribute to a redder appearance. These effects are borne out in our models which show that a zero-age O star
will have \mnh$-$\mnk~=$-$0.07, while a late-type nitrogen-rich WR (WNL)
star will have \mnh$-$\mnk~=+0.07 from the continuum alone, and \mnh$-$\mnk~=+0.13 from
the emission-lines and continuum. So, a WNL star will have a color that is about +0.20 magnitudes
redder than a zero-age O star. This agrees very well with the color trend seen in
Figure~4, after one removes the stars with \mnh$-$\mnk$>$1.85; these stars are subject to
extraordinary reddening, consistent with their location far from the cluster center and
the behavior of reddening as a function of increasing distance from the center of the cluster \citep{sto02}.

\subsection{Spectroscopy}
The spectra were reduced using IDL and IRAF routines. All target and calibration images were bias and
background subtracted, flat fielded, and corrected for bad pixels. The target images were
then transformed onto a rectified grid of data points spanning linear scales in
the spatial and wavelength directions using the
locations of stellar continuum sources and wavelength fiducials extracted from 
arc line images and continuum lamp plus etalon
images. The rectified images were then used as inputs to the aperture extraction
procedure. Finally, all 1D spectra were coadded and divided by the spectrum of
a telluric standard to produce final spectra. The following
gives a detailed description of this data reduction procedure.

A bias plus background image was produced from several images of a dark area of sky, observed with
the same instrument and detector parameters as the targets. Because the sky level was changing while 
these images were being obtained, we scaled the
sky components before combining them with a median filter.
The combined image was scaled in order to match the varying sky emission level
in the target images and then added to the bias image formed by taking the median of
several bias images. The resultant image was then subtracted from target images, thus 
subtracting properly scaled bias structure and background. This operation also removes dark current,
although the subtraction will be perfect only in the case that the scaling factor for
the varying sky level
is exactly one. In other cases, a small residual in dark current will remain, although
the amount of this residual will typically be less than 1 count (5 e$^{-}$).

Target images were divided by a normalized flat image, with bias structure and dark current first removed.  We 
then removed deviant pixels from these images with a two-pass
procedure. First, the median in a 5-by-5 pixel box surrounding each
pixel was subtracted from every pixel in order to form an image with
the low-spatial-frequency information removed. If
the absolute value of a pixel in this difference image was larger than five
times the deviation in the nearby pixels, then its value was replaced by the
median data value in the box. The deviation in nearby pixels is   
defined as the median of the absolute values of those pixels in the
difference image. On the second pass, isolated bad pixels were flagged
and replaced if both the following were true: they were higher or lower
than both immediate neighbors in the dispersion direction, and their   
value deviated by more than 10 times the square root of the average of
those two neighbors. Isolated bad pixel values are replaced by   
the average values of their neighbors.

We rectified the target images by mapping a set of continuum traces and wavelength
fiducials in the reduced images onto a set  
of grid points. This dewarping in both spatial and spectral directions 
is done simultaneously and requires knowledge of the wavelengths and
positions of several spectral and spatial features; this information was
extracted using arc, sky, or etalon
lines. Typically, we identified 15 to 20 lines in each echelle order for this purpose. Two stellar 
continuum spectra and two flat-field edges traced across the length of 
the dispersion direction were used to define spatial warping. 
Rectification was done separately for each echelle order.
 
Unfortunately, there is a lack of naturally occurring, and regularly spaced, wavelength fiducials (absorption
or emission lines) produced by the night sky or arc lamps. Because of this, we had
to use a three-stage process for determining the relationship between column number
and wavelength: rectify the images using the arc and sky lines, measure the etalon line wavelengths in 
the rectified version of the etalon image and obtain a solution to the etalon equation, and use 
the analytically determined wavelengths to produce a better
rectification matrix. This approach gave results that were
repeatable (to within an rms of $\sim$1~\kms) \citep{fig02}. This three-stage process
is described in detail below.
  
In the first stage, we chose an order containing many (15 to 20) arc 
and sky lines so that wavelengths in the rectified images would be
fairly well determined. The last few significant figures
of some of the arc-line wavelengths, and all of the OH lines, out to
seven significant figures, were derived from lists available from the 
National Institute of Standards and Technology (NIST) Atomic Spectra
Database\footnote{http://physics.nist.gov/PhysRefData/ASD1/nist-atomic-spectra.html}.

The process for fitting spectral lines was as follows. Each arc or sky line was    
divided into 10 to 20 sections along the length of the line. Rows 
from each section were averaged together, and 
the location of the peak was found by centroiding. Points along the line obtained from
the centroiding were typically fitted with a 3rd-order polynomial.
Outliers were clipped, and a new fit performed. We used a similar approach
to fit the stellar and flat-field edge spatial traces.
 
We then used the arrays of coefficients from spatial and spectral fits 
to produce a mapping between points along the spectral features in
the warped frame, and points in the dewarped frame. A two-dimensional 
second or third order polynomial was then fitted to these points, resulting
in a list of transformation coefficients, the ``rectification
matrix.''
 
In the second stage, after dewarping the etalon image with the
rectification matrix produced from the arc and sky lines, we measured 
the wavelength of each etalon line using SPLOT in IRAF. This provided 
preliminary estimated wavelengths for each
etalon line. Exact etalon wavelengths were given by the etalon equation. 
Solutions to the etalon equation were found using the constraint that  
features must have integer order numbers that decrease sequentially   
toward longer wavelengths.  The  
estimated wavelengths of the 14 to 16 lines were used to produce a
series of etalon equations that we simultaneously solved 
by finding the thickness and order numbers giving the least overall variance 
from the measured wavelengths.   
Once these parameters were determined, exact wavelengths could be
calculated for each etalon line.
 
In stage three, the exact etalon wavelengths were used to
determine a new rectification matrix by tracing these
features in the rectified etalon image. Together with the same spatial
information used to produce the  first stage rectification matrix, a   
new rectification matrix was produced.  The improved quality of the etalon  
lines allowed for a higher order (fourth or fifth order) polynomial fit
to the etalon mapping between warped and dewarped points. The new
matrices were applied to the appropriate spectral orders of target  
images.

Three images of the telluric standard (``Q3'') were taken using the
same setup as was used to obtain the target images. We moved the telescope along
the slit length direction between exposures so that the spectra were
imaged onto different rows of the detector. We reduced these images using the
sames procedures used for reducing the target images. 
 
Finally, APALL (IRAF) was used to extract spectra in manually chosen apertures. 
The resultant 1D spectra were then coadded in the case that a single object was
observed in multiple slit positionings. Before coadding, we shifted all spectra
to a common slit position by cross-correlating the telluric absorption features
and shifting the spectra. The coadded spectra of all the stars in Table~3 for which
we have spectra are shown in Figure~5; note that
these spectra have been smoothed using an 11-pixel (44~\kms) square boxcar
function for display purposes. The fluxes were not reddening-corrected, so the
spectral shape (roughly flat) is indicative of a hot star observed through a
large amount of extinction. 

\section{Results}
We use the data in this paper to form a census of stellar types for massive stars in the cluster,
estimate physical properties of the stars, determine the dynamical state of the cluster members, 
and assess the impact of the cluster on its environment.

\subsection{Spectral Types}
The spectra for the most massive stars in our data set are relatively similar, although the features have
smaller equivalent-widths 
for fainter stars. The brightest stars generally have spectra with a weak
feature near 2.058~\micron\ (\ion{He}{1}), weak emission near 2.104~\micron\ (\ion{N}{3}) and 
2.112/113~\micron\ (\ion{He}{1}), broad
and strong emission near 2.166~\micron\ (\ion{He}{1}, \ion{H}{1}), a weak line 
near 2.189~\micron\ in P-Cygni profile (\ion{He}{2}) in some cases, and
weak lines at 2.115/2.24/2.25~\micron\ (\ion{N}{3}), where the primary contributors to the
emission line fluxes are due to transitions of species listed in parentheses. 
There are indications of absorption at 2.058~\micron\ in most of the first ten stars, and
in some cases, it is strongly in a P-cygni profile (\#3 and \#8). In both these cases, the
absorption appears to have a ``double-bottom.'' This line and the 2.112/113~\micron\ blend
are narrow in the spectra of \#10, \#13, \#15, and \#29. 

From these spectra alone, we might assign spectral types of WNL (WN7$-$WN9) \citep{fig97} or O~If$^+$ \citep{han96}
for the brightest stars, just as have been assigned in \citet{nag95} and \citet{cot96}; 
the degeneracy in the classification of stars of these spectral types was noted by 
\citet{cont1995}. Figure~6 compares an average spectrum of the Arches stars with those of WNL stars \citep{fig97}
and O~If$^+$ stars \citep{han96}. We can see that the Arches stars have 
\ion{N}{3} emission at 2.104 \micron, 2.24 \micron, and 2.25 \micron, just like
that seen in the spectra of WN7$-$8 stars, while the
spectra of the O~If$^+$ stars do not have those lines in emission (or they are very weak). 
\citet{fig97} argued that \ion{N}{3} lines in the {\it K}-band might be used to distinguish Wolf-Rayet (WR) from O~If$^+$ stars
on both observational and theoretical grounds. In addition, \citet{fig97} showed
that WN stars separate by subtype as a function of the relative line strengths in certain lines
of their {\it K}-band spectra, i.e.\ W$_{2.189\, \mu m}$/W$_{2.166\, \mu m}$ or W$_{2.189\, \mu m}$/W$_{2.11\, \mu m}$.
The relevant values for spectra of the most massive Arches stars are consistent with classifications of WNL. In addition,
the equivalent-widths measured for the emission lines are in the range of those measured for WNL
stars in \citet{fig97}, but not for O~If$^+$ stars.
From all these measures, we conclude that the Arches stars are WN9 types, with an uncertainty of
$\pm$ one subtype. Later in this paper,
we use wind/atmosphere models to show that the estimated nitrogen, carbon, and helium abundances 
verify WNL classifications.

One objection to the WNL classification might be that it is improbable that all
the WR stars in the cluster be in the exact same evolutionary phase, i.e.\ WNL. Actually,
such a situation is predicted by \citet{mey95}. Indeed, our new observations show
that there are no WR stars in the cluster other than the WNL types, given
that our narrow-band observations would easily
detect the strong emission lines from all WR stars, such as the WNE star in the
Quintuplet cluster \citep{fig95b}, or the carbon-rich WR (WC) stars seen near 
the Pistol Star \citep{fig99b}. The lack of WC stars suggests an age less than 3.5~\Myr\ \citep{mey95},
consistent with the WNL classification for brightest stars in the Arches cluster.

We have identified two groups among the most massive stars with spectral types distinctly different
than the WN9 spectral type.  The group of stars from \#10 to
\#30 are O~If$^+$ types, and fainter stars are O main sequence stars. 

\subsection{Identification of the Main Sequence in the Galactic Center}
The fainter stars in Table~3 (ID$>$30) appear to have little to no excess emission at 1.87~\micron, 
consistent with the fact that their spectra appear to be
relatively flat. Indeed, some even show absorption at 2.058~\micron, 2.112~\micron, and 2.166~\micron,
of a few angstroms equivalent width. In the case of star \#68, the spectrum (Br$\gamma$ absorption), apparent magnitude, and 
extinction, suggest a luminosity and temperature consistent with an initial mass of about 60~\Msun, and its presence on
the main sequence, given an age of 2.5~\Myr\ and the Geneva models. The star is likely to be a late-O
giant or supergiant, albeit still burning hydrogen in its core, and thus its
classification on the main sequence; note that it is too bright to be a dwarf. There are several other similar stars that
have relatively featureless spectra, consistent with the spectra of O
main sequence stars. This result is a significant spectroscopic 
verification of the claim in \citet{fig99a} that the main sequence is clearly visible
in the broadband NICMOS data; also, note that \citet{ser99} identified likely O stars with spectra that
appeared to be featureless at the resolution of their observations.
While a possible identification 
of the main sequence in the Galactic Center has previously been claimed \citep{eck99,fig00a}, the result in this
paper is the first spectroscopic identification of single stars that are unambiguously
on the main sequence, as observed by their narrow absorption lines.

\subsection{Wind/atmosphere Modelling}
To model the massive Arches stars and estimate their physical parameters,
 we have used the iterative, non-LTE line blanketing method presented
by \citet{hil98}. The code solves the radiative transfer equation in the co-moving frame
 for the expanding atmospheres of early-type stars
in spherical geometry, subject to the constraints of statistical
and radiative equilibrium.
Steady state is assumed, and the density structure is set by the
mass-loss rate and the velocity field via the equation of continuity.
We allow for the presence of clumping via a clumping law characterized
by a volume filling factor f$(r)$, so that the ``smooth'' mass-loss rate, 
\Mdot$_S$, is related to the ``clumped'' mass-loss rate, \Mdot$_C$, through 
\Mdot$_S$=\Mdot$_C$/f$^{1/2}$. The velocity law \citep{hil89} is
characterized by an  isothermal effective scale height in the
inner atmosphere, and becomes  a $\beta$ law in the wind.
The  model is then prescribed by the stellar radius, \Rstar,
the stellar luminosity, \Lstar, the mass-loss rate \Mdot,
the velocity field, $v(r)$, the volume filling factor, f, 
and the abundances of the elements considered.
\citet{hil98,hil99} present a detailed
discussion of the code. For the present analysis we have assumed the atmosphere to be composed of
H, He, C, N, Mg, Si and Fe.

We created a grid of models within the parameter domain of interest. It was
bounded by 4.4$<$log(\Teff)$<$4.6, 
5.0$\leq$log(\Lstar/\Lsun)$\leq$6.5, $-$5.7$\leq$log(\Mdot/\Msunyr)$\leq$$-$3.8 , 0.25$\leq$H/He$\leq$10, and 1$\leq$Z/\Zsun$\leq$2. 
This grid was used as a starting point to perform detailed
analyses of the objects, for which the grid parameters were fine tuned, and
other stellar properties such as the velocity field and clumping law were
relaxed. The observational constraints were set by the NIRSPEC {\it K}-band spectra,
and the NICMOS F187W equivalent-width and continuum (F110W, F160W and F205W) values.

We present preliminary results for stars 
\#8 and \#10.  A complete, detailed analysis of the whole sample will be discussed in \citet{naj02}.
Table~4 shows the stellar parameters derived for stars  \#8 and \#10, while
Figures~7a,b show the excellent fits of the models to the observed spectra.
All stellar parameters for the strong-line-emission object (\#8) displayed in
Table~4 should be regarded as rather firm with the exception of the
carbon abundance which should be considered as an upper limit.
For star \#10, however, no information can be obtained about
its terminal velocity and clumping factor. Further, only upper limits
are found again for the carbon abundance, while the uncertainty in the 
nitrogen abundance is rather high, up to a factor of 3, due to the extreme
sensitivity of the \ion{N}{3} lines to the transition region between photosphere and wind in this
parameter domain. Therefore we have assumed the star to have a terminal
velocity of 1000~\kms, which is representative of this class of objects, and
a filling factor of f~=~0.1 as derived for object \#8. To illustrate how the
assumption of different clumping factor
or \Vinf  values can affect the derived stellar
parameters, we also show in Table~4 two additional models which match
as well the observed spectra with modified  f (f=1, \#10b) and  terminal velocity
(\Vinf=1600 \kms, \#10c). Note that there is no simple 
\Mdot$_{C1}$/f$_1$$^{1/2}$ = \Mdot$_{C2}$/f$_2$$^{1/2}$ scaling \citep{her01}
nor \Mdot$_1$/\Vinf$_1$ =\Mdot$_2$/\Vinf$_2$ 
scaling between the models, and that other stellar parameters require readjustment.

Table~4 reveals that objects \#8 and \#10 have very similar luminosities,
temperatures, and ionizing-photon rates. However, their
wind densities (\Mdot) and abundances reveal different evolutionary phases
for these two objects. Object \#8 fits well with a WNL evolutionary stage.
Its wind density and helium abundance are very similar to those derived
by \citet{boh99} for WN9h stars. On the other hand,
object \#10 can be placed into a O~If$^+$ phase as its wind density is roughly 
an order of magnitude lower than that of object \#8 and the derived He abundance
and upper limits for nitrogen enrichment indicate an earlier evolutionary phase. Indeed,
its {\it K}-band spectrum is nearly identical to that of the O8~If$^+$ star
HD151804 \citep{han96}. Interestingly, both Arches objects have luminosities about one magnitude larger
than their counterparts in \citet{boh99}.
 
\subsection{Ionizing Flux}
Containing so many massive stars, the Arches cluster produces a large ionizing flux.
We estimate the total ionizing flux emitted by the 
cluster to be 4(10$^{51}$)~photons~s$^{-1}$, based on our wind/atmosphere model fits for the two
emission-line stars in Table~4. To estimate the total flux, 
we multiplied the estimated ionizing flux from \#8 by
10, that from \#10 by 20, and determined those of the remaining stars in Table~3 by 
applying equation 3 in \citet{cro98}; these factors reflect our choice of \#8 and \#10 as representatives
of the first 30 stars in the table.
The ionizing flux estimate is a bit higher than that in \citet{ser98}, $\approx$ a few 10$^{51}$~photons~s$^{-1}$,
after scaling that number for the fact that the cluster contains 50\% more O stars
than originally thought \citep{fig99a}. This amount of ionizing flux is consistent
with the Arches cluster being the ionizing source for the Thermal Arches Filaments \citep{lan01b}.

\subsection{Luminosity}
Using the same process as applied to estimate the cluster ionizing flux, we estimate a total
cluster luminosity of 10$^{7.8}$~\Lsun, or one of the most luminous clusters in the Galaxy. About 40\% of the total luminosity is
contributed by the 30 brightest stars in Table~3.

\subsection{Age}
As described in \citet{fig99b}, the absolute magnitudes and mix of spectral types are consistent with a cluster age
of ~2$\pm$1~\Myr. This was estimated using the colors and magnitudes of the stars, 
i.e.\ the colors give the extinction value, and the apparent magnitudes lead to absolute
{\it K}-band magnitudes, which are then compared to isochrones from the Geneva models \citep{mey94}. It is possible
for older stars to attain magnitudes as bright as the brightest stars in the cluster, but only at relatively cool temperatures, 
i.e.\ T$_{\rm eff}~<$~25kK. A new age constraint from the data in this paper is given by the
absence of WC or WNE stars in the cluster. This observational constraint, when combined with
the models of \citet{mey95}, gives
$\tau_{\rm age}~<~3.0~\Myr$ for the least limiting case (2$\times$\Mdot, Z=0.040) and
$\tau_{\rm age}~<~2.5~\Myr$ for the most limiting case (1$\times$\Mdot, Z=0.040). In addition,
the presence of WNL stars requires $\tau_{\rm age}~>~1.5~\Myr$ from these models. Finally, we
note the lack of relatively cool (B-type) supergiant emission-line stars, such as those found
in the central parsec \citep{kra95} and the Quintuplet cluster \citep{fig99b}. 
The lack of such stars in the Arches cluster implies $\tau_{\rm age}~<~4.0~\Myr$. Finally,
the detailed model for star \#8 suggests an age of 2.5~\Myr, at least for that star. We
combine all this evidence to suggest that the cluster age is 2.5~$\pm$0.5~\Myr, where the error
is dominated by our lack of information concerning metallicity.

\subsection{Velocities}
A velocity determination for the emission-line stars is complicated by several facts. First, the strongest
spectral features are blended emission lines, making it impractical to simply compare the measured wavelength centroid of a
``line'' to the expected vacuum wavelength. Because of this, one must cross-correlate the target spectra with
respect to a template spectrum composed of features that accurately represent relative strengths of the blended lines. 
Second, the emission lines are broad, so that 
small wavelength shifts will produce little change in the cross-correlated power when compared to a template spectrum.
Third, there are slight differences in the shapes of the emission lines between spectra of the
various stars, so a model blend for one spectrum might not faithfully reproduce the features in
another spectrum, at least not to the fidelity required for high precision velocity measurements.
Because of these difficulties, we approached the velocity estimates using two techniques. We smoothed
the spectra using a box-car filter with varying widths between 1 and 31 pixels (4 and 120~\kms), finding
little difference in the velocities as a function of these widths.

First, we estimated an absolute velocity for star \#8 by cross-correlating its spectrum with that of
our model spectrum. This method gave a velocity of +54~\kms\ (redshifted), in the heliocentric frame,
for the blend near 2.166~\micron. Our estimates using other lines are somewhat less than this value, as low as +20~\kms,
but those lines are weaker than the blend at 2.166~\micron, and thus produce larger velocity errors.

In the second method, we cross-correlated all the spectra against each other. The cross-correlations
were performed separately on three wavelength regions, the first containing the 2.104~\micron\ and
2.115~\micron\ features, the second containing the 2.166~\micron\ and 2.189~\micron\ features, and the third containing
the doublet at 2.25~\micron. This method was used to compute the standard deviation of relative velocities, 
allowing us to infer the mass enclosed within some orbital radius that represents an average
of the emission-line stars' orbital radii. Given that small differences in intrinsic 
blend morphology can affect the location of the maximum point of the cross-correlated power, we 
also repeated this approach using line centroids to compute velocity differences. Using both
approaches, we found a standard deviation of $\approx$22~\kms\ for a sample containing eight emission-line stars. 
This value represents an upper limit on the intrinsic dispersion,
given that the effects described above would tend to increase the estimated value over the
intrinsic velocity dispersion. We also found that the stars were redshifted by +41~\kms\ with
respect to star \#8. We therefore estimate a heliocentric ``cluster'' velocity of +95$\pm8$~\kms,
where the error is simply the standard deviation divided by the square root of eight; note that
this error neglects the systematic effects described above, so the true velocity might differ
from the estimate by significantly larger than the quoted error.

For a gravitationally bound and spherically symmetric cluster of mass M$_{\rm cluster}$, the virial theorem gives
M$_{\rm cluster}$ = 3$\sigma^2$R/G \citep{hof96}, where $\sigma$ is the one-dimensional velocity dispersion,
R is the appropriate radius, G is the gravitational constant, the velocities are assumed
to be isotropic, and all stars have equal mass. This simple formula can be compared to
the more general case where the cluster can be resolved into individual stars \citep{ill76}.
Using R~=~0.23~pc for the sample in question, we calculate M$_{\rm cluster}$~$<$~7(10$^4$)~\Msun,
or about five times greater than what would be expected from direct integration of 
the mass function in \citet{fig99a} over the area sampled by the stars used in the analysis. 
This high mass limit results from the systematic effects inherent in our radial velocity determinations, as described
above.

\section{Discussion}
In this section, we compare our measurements to those in previous papers and use measurements at other wavelengths 
to determine the physical parameters of the observed stars. Finally, we
discuss how the Arches cluster interacts with its local environment to create 
heating and ionization of a nearby cloud.

\subsection{Comparison to Previous Near-infrared Measurements}
Table~3 lists over 30 probable emission-line stars, albeit the faintest having relatively
weak emission lines; Figure~3a confirms that there are roughly this number of stars with reliable
emission-line excesses. This list contains over a factor of two increase in the number of
emission-line stars previously identified in the cluster \citep{blu01,nag95,cot96}.
The line and continuum fluxes presented here largely agree with earlier results \citep{nag95,cot96}. 
The \Fnp\ and \Fnc\ fluxes reported in this paper are similar to those reported in \citet{blu01}, 
after correcting for differences in the assumed zero points, the fact that we 
corrected for the difference in extinction at the two narrow-band wavelengths, and that we also
corrected for the intrinsic shape of the stellar continuum; in addition, our extinction estimates
are higher in many cases than those used in \citet{blu01}.

The spectra in this paper are consistent with the narrow-band photometry in \citet{nag95} and \citet{blu01} and
the spectra in \citet{cot96}, although our high-resolution spectroscopy shows that the photometry 
is significantly affected by blending of absorption and emission features in P-Cygni profiles. 

We confirm the discovery of a new bright emission-line star (\#5, B22, N9) near the southern edge 
of the cluster, reported in \citet{blu01}, and note that it is a counterpart 
of the x-ray source, ``AR8,'' in \citet{lan01a}. We also confirm that
star \#16 (B19) is an emission-line star, as suspected by \citet{blu01}. 
\citet{blu01} listed some additional candidate emission-line stars. 
We confirm that the following stars
from that list are, indeed, emission-line stars (their designations in parentheses):
\#15 (B8), \#27 (B16), \#17 (B29), \#10 (B30), \#10 (B20), \#13 (B31). 

\subsection{Comparison to X-ray Flux Measurements}
\citet{zad01} reported Chandra X-ray observations of a region
including the Arches cluster.  They detected three extended sources, one (A1)
near the center of the Arches cluster, another (A2) located to the North and
West of the center by about 7\arcsec\, and a third weaker source (A3) about 90\arcsec\ $\times$  60\arcsec\
in size underlying the first two.  The centroid of source A2 coincides within 1
arcsecond with an emission-line star, \#9 in Table~3.  The apparent spatial
coincidence of the X-ray sources and Arches cluster strongly suggests that
the X-ray sources are physically associated with the cluster.  Yusef-Zadeh
et al. estimate the total X-ray luminosity between 0.2 and 10 keV to be 3.3,
0.8 and 0.16 (10$^{35}$) ergs s$^{-1}$ for A1, A2 and A3, respectively.
They attribute the emission from A1 and A2 to either colliding winds in binary
systems or to the winds from single stars interacting with the collective wind
from the entire cluster.  The coincidence of source A2 with an emission-line source 
is very interesting in the context of the latter scenario.  A3, on the other
hand, has roughly the characteristics expected from shock-heated gas created
by the collisions of the multitude of 1000-\kms\ stellar winds
emanating from the stars in the rich, dense cluster \citep{oze97,can00}.  Because the X-ray sources are extended,
it is unlikely that they can be attributed to single X-ray binary systems.
However, the rough coincidence of source A1 with the core of the cluster
raises the possibility that it may be comprised of many relatively weak
stellar X-ray sources, binary or single, residing in the cluster core, and
unresolved spatially from each other.

\subsection{Comparison to Radio Flux Measurements}
From Table~3 we see that one of the objects analysed in this work, \#8, has
also been detected at 8.5~GHz \citep{lan01a}. Our derived mass-loss rate
is consistent with the observed radio flux (0.23~mJy) only if the outer
wind regions are unclumped. Such a behavior for the clumping law has been
suggested by \citet{nug98} from analysis of galactic WR stars. They found that
the  observed infrared to radio fluxes of WR stars are well reproduced by
a clumping law where the filling factor is unity close to the stellar surface, increasing
to a minimum at 5 to 10 R$_*$ and returning again to unity in the outer
wind where the radio flux forms.
Note, however, that the line fluxes of the
weaker lines like  Br$\gamma$ or \ion{He}{1} remain unaltered
with this new description of the clumping law, but the line fluxes of the
strongest lines, such as Pa$\alpha$, formed in the outer wind  can be
significantly reduced. From Table~4 we see that our models are fully
consistent with the observed equivalent-width of Pa$\alpha$.

We consider now the possible correlation between line fluxes
and radio-continuum flux analogous to the one discussed above for {\it K}-band fluxes
(see Figure~3a). In principle,
we also expect the near-infrared emission line strengths to scale with the
free-free emission
detected at radio wavelengths \citep{nug98,lei97}.  However, we do not find
such a correlation, as can be seen in Figures~8a,b. A similar result was
obtained by \citet{bie82} for a sample of eight WR stars.
We believe this apparent absence of correlation between
Pa$\alpha$ line-strength and radio flux is caused by both  observational
and  physical effects. The observational effect is related to
the fact that the radio measurements are picking up only the tip
of the iceberg, i.e., those stars with the densest winds of the cluster.
The physical effect is related to the fact that all three components
contributing to EW$_{\rm 1.87\micron}$ (\ion{H}{1}, \ion{He}{1}, and \ion{He}{2}) are
very sensitive to changes in temperature in the parameter domain appropriate to these objects. Further, both the line and continuum
fluxes depend strongly not only on the mass-loss rate but also on the
shape of the velocity field and the clumping law. Therefore, such strong
dependence of the Pa$\alpha$ line flux on several stellar parameters
introduces a large scatter in the expected line-strength vs radio-flux
relationship.

The radio fluxes of the most massive Arches stars are comparable to those of WNL stars, but not
to those of O~If$^+$ stars. 
The WN8 star WR105 \citep{van01} would emit 0.14 mJy at
the distance of the Arches cluster, comfortably within the range of fluxes measured for the
Arches stars. Similar values are reported for WR stars in \citet{bie82}.
On the other hand, HD 16691 (O4~If$^+$) emits 0.3~mJy at 4.9~GHz, according to \citet{wen95}, 
implying an expected flux of 1.7~$\mu$Jy
at the distance of the Arches cluster, assuming that the star has a parallax of 1.7~mas \citep{per97}.
The expected flux is two orders of magnitude below the flux levels of the brightest Arches stars 
\citep{lan01a}. No doubt, this difference is due to the relatively low mass-loss rate for HD 16691, about 1/20 of
that of the bright Arches stars. A similar trend can be seen in Figure~6 where the emission
lines in the spectra of HD 16691 are shown to be much weaker than those in the spectra of the Arches
emission-line stars. Again, weak winds produce weak emission lines and weak free-free
emission.

Finally, we report several additional radio sources having emission-line star
counterparts. They were found by comparing the radio continuum contour
plot in \citet{lan01a} with the difference image in Figure~2; they are marked by
squares in this figure. We have designated the four 
near-infrared counterparts to these newly identified radio sources in Table~3.

\subsection{Evolutionary Status of the Massive Stars in the Arches Cluster}
The emission-line stars appear to contain significant amounts of hydrogen, while
also exhibiting considerable helium content. We believe that this can be explained
by the most massive stellar models in \citet{mey94}. For the brightest 10 or so
stars in Table~3, the observations can be fit
by these models for \Minit$\gtrsim$120~\Msun\ stars that have evolved to cool temperatures while
retaining hydrogen. In particular, star \#8 can be fit by a \Minit$\sim$120~\Msun\ star with solar abundance, 
standard mass-loss rates, age of 2.4~\Myr\ to 2.5~\Myr, and present-day mass of 72~\Msun\ to 76~\Msun\ \citep{sch92}.

\subsection{Relation to the Nearby Molecular Cloud (M0.10+0.03)}
It appears that the Arches cluster heats and ionizes the surface of M0.10+0.03, the nearby
molecular cloud \citep{ser87,bro84}, given that the cluster can easily provide
the necessary flux to account for the infrared emission and recombination-line flux
from the cloud. 

The relative heliocentric velocity between the cluster stars (+95$\pm8$~\kms) and the ionized gas on the surface of the
cloud ($-$20 to $-$50~\kms) suggests that the physical association is accidental and that the
cluster stars are ionizing the surface of the cloud. This difference in velocity is reminiscient of that
observed between the Quintuplet (+130~\kms) \citep{fig95a,fig99a} and the Sickle cloud,
M0.10+0.03 (+30~\kms) \citep{lan97}.
In both cases, it appears that young clusters happen to lie near molecular clouds whose surfaces are ionized
by the photons from the hot stars in the clusters. The following shows that the ionizing flux
and energy required to heat the cloud can be provided by the Arches cluster. Note that a
differential velocity of 100~\kms\ would produce a relative drift of 100~pc in 1~\Myr, a distance
that would bring a cluster within the vicinity of a few clouds, given the spatial 
distribution of clouds in the central few hundred parsecs.

\subsubsection{Ionization and Heating}
Even before the discovery of the Arches cluster, \citet{ser87}, \citet{gen90}, and \citet{miz94} 
suggested that the Thermal Arched Filaments are photoionized by nearby hot stars. After the discovery
of the cluster, many authors considered the possibility that the cluster is ionizing the cloud.
One problem with this idea is the fact that the filaments are very large, and have 
roughly constant surface brightness and excitation conditions \citep{eri91,col96}, indicating
that the ionizing source is either evenly distributed over many parsecs or is
relatively far away. Given the new ionizing flux estimates in this paper and in \citep{ser98}, the
cluster would produce enough flux to account for the filaments, even if 20~pc away, far 
enough to allow for the even illumination that is observed. Indeed, 
\citet{tim96} predicted this result, and \citet{lan01b} present a detailed analysis that confirms
it.

\citet{cot96} give estimates for the total ionizing flux of 2$-$5(10$^{50}$) photons s$^{-1}$, depending
on whether one models the spectral energy distributions for the emission-line stars
with \citet{kur79} atmospheres or blackbody functions; however, this estimate 
includes only flux from the dozen or so emission-line stars that were known at the time. Nonetheless,
the results of this paper suggest that the cluster ionizes the cloud.

The heating in the cloud produces an infrared luminosity of 10$^7$~\Lsun\ \citep{mor95}.
Assuming a covering fraction of $\approx$10\%, we find that the Arches cluster can
deliver about this much luminosity, within a factor of two. 

\subsubsection{Location along the line of sight}
Given the Br$\gamma$ flux from the filaments measured 
by \citet{fig95a}, we know that the line emission is extincted by about \AK$\sim$3. This
implies that the filaments are on the near side of the cloud, since such an extinction corresponds only
to the typical foreground extinction to the Galactic Center, and precludes any substantial additonal extinction. This information leads 
to the conclusion that the Arches cluster is on the near side of, and is approaching, the
wayward molecular cloud that is moving in opposition to the bulk motion of stars and gas around
the Galactic Center \citep{mcg89}, consistent with the geometry described in \citet{lan01b}.

\subsection{Dynamical Evolution and Uniqueness of the Arches and Quintuplet Clusters}
The temporal coincidence of the star formation events that produced the massive clusters
in the Galactic Center, and the lack of older red supergiants, suggest that the
Galactic Center has been host to a recent burst of star formation. 
 \citet{kim99} and \citet{kim00} predicted that compact young clusters in the Galactic
Center would evaporate on short timescales, i.e.\ a few \Myr. 
 \citet{zwa01a} argue that other clusters 
similar to, yet somewhat older than, the Arches and Quintuplet
clusters exist in the central 100~pc. This argument is based upon a dynamical analysis
which predicts that such clusters 
evaporate after 55~\Myr, and further that the clusters' projected surface number density
in stars drops below the limit of detectability in a few \Myr.
The statement that members of dispersed clusters could have gone undetected is incorrect.
Such stars would easily be detectable for their extreme brightness, i.e. there would
be hundreds of stars as bright as IRS~7 (in the central parsec) strewn
about the central 100~pc for each Arches/Quintuplet-like cluster between the age of
5 and 30~\Myr. Given the claim in Portegies-Zwart et al.\ about the expected number of ``hidden''
young clusters in the central hundred parsecs, we would expect to see of order ten
thousand red supergiants in this region. Only a few are seen, as demonstrated by 
surveys for such stars \citep{fig95a,cot95}. 

Portegies-Zwart et al.\ suggest 
that clusters could be ``hiding'' near bright stars due to limitations in 
dynamic range, but these arguments are specious, since the dynamic range 
of array-based detections are obviously not limited by the digitization 
of a single read if multiple coadds are used (e.g.\ \citet{fig99a} reach a 
dynamic range of over 10$^5$). Thus in 
agreement with Kim et al., we conclude that the clusters must disperse 
rapidly

\subsection{Comparison to NGC~3603 and R136 in 30 Dor}
The Arches cluster is similar in age and content to NGC~3603 and R136 in 30 Dor, and is
surrounded by a giant \ion{H}{2} region as is R136. 
In contrast to these clusters, we do not see WN5h or WN6h stars \citep{cro98}, suggesting
that the Arches cluster is older. This is consistent with our age
determination, as suggested by other means described earlier.

While our spectra exhibit no primary diagnostic lines
to estimate metallicity, we may use our estimates for the helium and nitrogen abundances
in object \#8 in conjunction with the evolutionary model for 120~\Msun\ to infer metallicity.
For Z(He)=0.7, the models predict the star to have already reached
its maximum nitrogen surface mass fraction. Hence, we may compare our
derived nitrogen mass fraction, Z(N)=0.016, with the evolutionary models values at different
metallicities \citep{sch92}. We see that this value is
met for solar metallicity. A more detailed analysis of the metallicity of the 
Arches stars will be presented in \citet{naj02}.

\acknowledgements
We thank Nolan Walborn and Nino Panagia for critical readings of the paper.
We thank John Hillier for providing his code.
  F. N. acknowledges DGYCIT grants ESP98-1351 and PANAYA2000-1784.
We acknowledge the work of: Maryanne Angliongto, Oddvar
Bendiksen, George Brims, Leah Buchholz, John Canfield, Kim Chin, Jonah
Hare, Fred Lacayanga, Samuel B. Larson, Tim Liu, Nick Magnone, Gunnar
Skulason, Michael Spencer, Jason Weiss and Woon Wong.  

\clearpage
\begin{deluxetable}{rcrrrcr}
\tabletypesize{\scriptsize}
\tablewidth{0pt}
\tablecaption{Log of Observations}
\tablehead{
\colhead{Type} &
\colhead{Filter\tablenotemark{a}} &
\colhead{$\lambda_{\rm center}$} &
\colhead{PHOTFNU/ZP(Vega)\tablenotemark{b}} &
\colhead{Ap.\ Corr.} &
\colhead{Integ.} &
\colhead{Date}  \\
\colhead{} &
\colhead{} &
\colhead{\micron} &
\colhead{sec. DN$^{-1}$} &
\colhead{3~pix. to $\inf$} &
\colhead{} &
\colhead{}
}
\startdata
Spectroscopy\tablenotemark{c} & NIRSPEC-6       & {\it K}-band & \nodata & \nodata & 150 s. & 3 July 1999 \\
Imaging &  F110W & 1.10   & 9.61E-10   & 1.35 & 256 s. & 13 September 1997 \\
Imaging &  F160W & 1.60         & 1.86E-09   & 1.67 & 256 s. & 13 September 1997  \\
Imaging &  F205W & 2.05   & 1.64E-09   & 1.81 & 256 s. & 13 September 1997 \\
Imaging &  F187N & 1.874        & 4.95E-08    & 1.75 & 256 s. & 13 September 1997  \\
Imaging &  F190N & 1.900        &  5.36E-08   &  1.76 & 256 s. & 13 September 1997 \\
\enddata
\tablenotetext{a}{NIRSPEC-6 has half-power points of 1.85~\micron\ and
2.62~\micron\ \citep{fig00b}. Because the orders are longer than the width of the
detector, the spectra are not contiguous in wavelength.}
\tablenotetext{b}{Multiplying PHOTFNU/ZP(Vega) by the observed count rate gives the ratio
of the object's flux with respect to Vega.
Values for F110W, F160W, and F205W are from the HST Data Handbook \citep{key97}. Values for
F187N and F190N are from Marcia Rieke (priv. communication). The zero points in \citet{blu01} were
also provided by Marcia Rieke; however, the values used in the table above were
more recently provided.}
\tablenotetext{c}{All spectroscopy images were obtained with the slit
positioned approximately north-south using the multiple correlated read mode (``Fowler'' sampling) with 16 reads
at the beginning and end of each integration. The resolution was 23,300,
($\lambda$/$\Delta\lambda_{\rm FWHM}$, where
$\Delta\lambda_{\rm FWHM}$ is the full-width at half maximum of unresolved
arc lamp lines). The slit size was 0$\farcs43\times 24\arcsec$.}
\end{deluxetable}

\clearpage
\begin{deluxetable}{crccrcc}
\small
\tablewidth{0pt}
\tablecaption{Wavelength Coverage in Spectra}
\tablehead{
\colhead{Echelle} & \colhead{} & \colhead{$\lambda_{\rm min}$} & \colhead{$\lambda_{\rm max}$} \\
\colhead{Order} & \colhead{} & \colhead{\micron} & \colhead{\micron} 
 }
\startdata
33  & &  2.281 & 2.315   \\
34  & &  2.214 & 2.248   \\
35  & &  2.152 & 2.184   \\
36  & &  2.092 & 2.124   \\
37  & &  2.036 & 2.067   \\
38  & &  1.983 & 2.013   \\
\enddata
\end{deluxetable}

\clearpage
\begin{deluxetable}{rlrrrrrrrrr}
\tabletypesize{\scriptsize}
\tablewidth{0pt}
\tablecaption{Massive Stars in Arches Cluster}
\tablehead{
\colhead{ID\tablenotemark{a}} &
\colhead{Desig./Ref.\tablenotemark{b}} &
\colhead{$\Delta$RA\tablenotemark{c}} &
\colhead{$\Delta$DEC\tablenotemark{c}} &
\colhead{m$_{\rm F110W}$} &
\colhead{m$_{\rm F160W}$} &
\colhead{m$_{\rm F205W}$} &
\colhead{M$_{\rm K}$\tablenotemark{d}} &
\colhead{M$_{\rm init}$\tablenotemark{e}} &
\colhead{EW\tablenotemark{f}} &
\colhead{F$_{\rm F190N}$} \\
\colhead{} &
\colhead{} &
\colhead{$\arcsec$} &
\colhead{$\arcsec$} &
\colhead{mag.} &
\colhead{mag.} &
\colhead{mag.} &
\colhead{mag.} &
\colhead{\Msun} &
\colhead{$\AA$} &
\colhead{$ergs~cm^{-2} s^{-1} \AA^{-1}$} \\
\colhead{(1)} &
\colhead{(2)} &
\colhead{(3)} &
\colhead{(4)} &
\colhead{(5)} &
\colhead{(6)} &
\colhead{(7)} &
\colhead{(8)} &
\colhead{(9)} &
\colhead{(10)} &
\colhead{(11)} 
}
\startdata
1 & N4 C9 AR3 B28 &      0\farcs00 &     0\farcs00 & 16.30 & 12.33 & 10.45 & $-$8.0 & $>$120 &128.2 & 3.00e-015 \\
2 & N1 C13 AR17 B34 &   $-$6\farcs75 &  $-$3\farcs53 & 17.84 & 13.39 & 11.18 & $-$7.9 & $>$120 &152.3 & 1.39e-015 \\
3 & N14 C11 AR7 B3 &      8\farcs20 &  $-$4\farcs13 & 16.06 & 12.28 & 10.46 & $-$7.7 & $>$120 &218.2 & 2.88e-015 \\
4 & N11 C2 AR5 B17 &      4\farcs83 &     4\farcs66 & 15.63 & 12.12 & 10.37 & $-$7.6 & $>$120 &230.5 & 3.18e-015 \\
5 & N9  AR8 B22 &      3\farcs29 &  $-$9\farcs64 & 16.69 & 12.80 & 10.86 & $-$7.6 & $>$120 &199.3 & 2.02e-015 \\
6 & N8 C8 AR1 B23 &      2\farcs87 &  $-$0\farcs03 & 15.75 & 12.05 & 10.37 & $-$7.6 & $>$120 &166.3 & 3.83e-015 \\
7 & N10 C5 AR4 B21 &      3\farcs53 &     2\farcs73 & 15.74 & 12.16 & 10.48 & $-$7.5 & $>$120 &150.2 & 3.25e-015 \\
8 & N7 C6 AR2 B24 &      2\farcs46 &     1\farcs01 & 16.31 & 12.54 & 10.76 & $-$7.3 & $>$120 &206.0 & 2.55e-015 \\
9 & N5 C1   &      0\farcs80 &    10\farcs50 & 16.10 & 12.44 & 10.77 & $-$7.3 & $>$120 &106.7 & 2.28e-015 \\
10 &    B30 &   $-$1\farcs83 &  $-$4\farcs25 & 17.37 & 13.35 & 11.46 & $-$7.1 & $>$120 &51.8 & 1.28e-015 \\
11 & \nodata &   $-$1\farcs03 &    14\farcs41 & 17.02 & 12.72 & 10.92 & $-$7.1 & $>$120 &\nodata & \nodata \\
12 & N6 C3 AR16 B25 &      1\farcs01 &     4\farcs98 & 16.40 & 12.67 & 10.99 & $-$7.0 & $>$120 &162.7 & 1.72e-015 \\
13 &    B31 &   $-$2\farcs08 &  $-$1\farcs39 & 17.59 & 13.63 & 11.74 & $-$6.9 & 116.9 &18.2 & 9.39e-016 \\
14 &    B12 &      6\farcs24 &  $-$0\farcs32 & 16.38 & 12.84 & 11.22 & $-$6.7 & 106.3 &124.9 & 1.59e-015 \\
15 & N12 CB  B8 &      7\farcs24 &     5\farcs67 & 16.12 & 12.78 & 11.27 & $-$6.5 & 99.7 &41.2 & 1.69e-015 \\
16 &    B19 &      4\farcs22 &     1\farcs59 & 16.62 & 13.01 & 11.40 & $-$6.5 & 99.5 &97.5 & 1.39e-015 \\
17 &    B29 &   $-$0\farcs89 &  $-$4\farcs90 & 18.13 & 14.05 & 12.15 & $-$6.5 & 97.6 &47.1 & 6.69e-016 \\
18 &   AR9 B20 &      3\farcs58 &     4\farcs34 & 16.70 & 13.17 & 11.63 & $-$6.2 & 92.0 &17.6 & 1.10e-015 \\
19 &   AR6  &   $-$5\farcs81 &  $-$3\farcs72 & 18.89 & 14.58 & 12.60 & $-$6.2 & 92.0 &0.6 & 4.03e-016 \\
20 & \nodata &      2\farcs90 &     2\farcs58 & 17.49 & 13.88 & 12.16 & $-$6.1 & 89.1 &1.6 & 5.89e-016 \\
21 &    B7 &      7\farcs36 &     2\farcs65 & 16.85 & 13.29 & 11.77 & $-$6.1 & 87.8 &17.2 & 1.00e-015 \\
22 &    B27 &      0\farcs24 &     5\farcs55 & 17.46 & 13.65 & 12.02 & $-$6.1 & 87.7 &11.4 & 7.25e-016 \\
23 &    B2 &     12\farcs50 &  $-$1\farcs08 & 17.56 & 13.82 & 12.19 & $-$5.9 & 70.3 &11.4 & 6.40e-016 \\
24 & \nodata &   $-$1\farcs42 &     1\farcs55 & 18.27 & 14.40 & 12.61 & $-$5.8 & 66.7 &1.4 & 4.34e-016 \\
25 & \nodata &   $-$3\farcs26 &  $-$4\farcs30 & 19.42 & 15.05 & 13.05 & $-$5.8 & 66.1 &10.9 & 2.81e-016 \\
26 &    B18 &      4\farcs60 &  $-$1\farcs27 & 17.70 & 13.98 & 12.34 & $-$5.8 & 64.9 &$-$0.7 & 6.10e-016 \\
27 &    B16 &      5\farcs31 &     2\farcs74 & 17.13 & 13.49 & 12.01 & $-$5.7 & 63.0 &22.1 & 8.58e-016 \\
28 &    B14 &      5\farcs77 &     0\farcs55 & 17.26 & 13.69 & 12.17 & $-$5.7 & 61.7 &6.7 & 6.83e-016 \\
29 &    B9 &      7\farcs08 &     4\farcs62 & 17.23 & 13.81 & 12.26 & $-$5.7 & 61.0 &3.5 & 6.49e-016 \\
30 & \nodata &      0\farcs20 &     3\farcs66 & 17.87 & 14.16 & 12.53 & $-$5.6 & 59.0 &1.5 & 4.67e-016 \\
31 & \nodata &      2\farcs87 &     2\farcs60 & 17.70 & 13.97 & 12.41 & $-$5.5 & 58.7 &1.6 & 5.89e-016 \\
32 &    B15 &      5\farcs53 &     2\farcs41 & 17.59 & 13.96 & 12.42 & $-$5.5 & 57.6 &8.7 & 5.52e-016 \\
33 &    B13 &      6\farcs08 &     2\farcs36 & 17.53 & 13.95 & 12.42 & $-$5.4 & 57.0 &23.2 & 5.40e-016 \\
34 &    B5 &      7\farcs93 &     1\farcs22 & 17.67 & 14.03 & 12.49 & $-$5.4 & 56.6 &6.3 & 5.31e-016 \\
35 &    B10 &      6\farcs51 &     5\farcs26 & 17.29 & 13.84 & 12.37 & $-$5.4 & 56.1 &8.4 & 6.10e-016 \\
36 & \nodata &   $-$6\farcs19 &    14\farcs87 & 19.03 & 14.53 & 12.60 & $-$5.4 & 56.0 &\nodata & \nodata \\
37 & N10 C5 AR4 B21 &      3\farcs54 &     2\farcs99 & 99.00 & 14.55 & 12.63 & $-$5.3 & 55.5 &46.3 & 3.14e-017 \\
38 &    B11 &      6\farcs51 &     2\farcs28 & 17.51 & 13.85 & 12.38 & $-$5.3 & 55.3 &38.6 & 5.64e-016 \\
39 & \nodata &     11\farcs91 & $-$13\farcs94 & 18.10 & 14.37 & 12.65 & $-$5.3 & 55.2 &\nodata & \nodata \\
40 & \nodata &      5\farcs59 &     3\farcs93 & 17.78 & 14.24 & 12.67 & $-$5.3 & 54.7 &10.7 & 4.57e-016 \\
41 & \nodata &      2\farcs22 &  $-$4\farcs66 & 19.53 & 15.47 & 13.53 & $-$5.2 & 53.5 &6.6 & 1.85e-016 \\
42 & \nodata &      1\farcs37 &     1\farcs52 & 18.02 & 14.41 & 12.82 & $-$5.2 & 53.0 &2.2 & 3.71e-016 \\
43 & \nodata &      3\farcs76 &     3\farcs59 & 18.45 & 14.69 & 13.04 & $-$5.1 & 51.9 &$-$3.7 & 2.64e-016 \\
44 & \nodata &      5\farcs80 &  $-$2\farcs83 & 18.25 & 14.47 & 12.88 & $-$5.1 & 51.9 &1.2 & 3.59e-016 \\
45 & \nodata &      3\farcs11 &  $-$1\farcs49 & 18.47 & 14.64 & 13.01 & $-$5.1 & 50.9 &6.8 & 3.35e-016 \\
46 & \nodata &     11\farcs18 &  $-$5\farcs42 & 19.03 & 14.69 & 12.90 & $-$5.1 & 50.6 &\nodata & \nodata \\
47 & \nodata &      6\farcs49 &  $-$2\farcs52 & 18.01 & 14.46 & 12.90 & $-$5.0 & 50.2 &5.0 & 3.58e-016 \\
48 & \nodata &      4\farcs77 &  $-$4\farcs20 & 18.73 & 15.00 & 13.28 & $-$5.0 & 50.2 &$-$3.2 & 2.39e-016 \\
49 & \nodata &   $-$1\farcs74 &    14\farcs97 & 18.36 & 14.59 & 12.94 & $-$5.0 & 49.9 &\nodata & \nodata \\
50 & \nodata &   $-$1\farcs65 &  $-$3\farcs41 & 19.57 & 15.38 & 13.53 & $-$5.0 & 49.9 &2.1 & 2.01e-016 \\
51 & \nodata &     10\farcs80 &  $-$1\farcs62 & 19.82 & 14.99 & 12.94 & $-$5.0 & 49.9 &$-$7.8 & 2.94e-016 \\
52 & \nodata &     12\farcs43 & $-$10\farcs55 & 18.59 & 14.67 & 12.94 & $-$5.0 & 49.9 &\nodata & \nodata \\
53 & \nodata &      5\farcs96 &  $-$2\farcs25 & 18.25 & 14.51 & 12.94 & $-$5.0 & 49.7 &6.4 & 3.53e-016 \\
54 & \nodata &      0\farcs14 &     6\farcs95 & 18.77 & 14.78 & 13.02 & $-$5.0 & 48.4 &$-$5.4 & 2.75e-016 \\
55 & \nodata &      6\farcs93 &     0\farcs15 & 18.13 & 14.57 & 13.03 & $-$4.9 & 48.1 &$-$10.4 & 3.13e-016 \\
56 & \nodata &      4\farcs35 &     0\farcs57 & 18.15 & 14.54 & 13.03 & $-$4.9 & 48.1 &$-$11.4 & 3.33e-016 \\
57 &    B11 &      6\farcs67 &     2\farcs45 & 18.01 & 14.53 & 13.04 & $-$4.9 & 47.9 &\nodata & \nodata \\
58 & \nodata &   $-$3\farcs84 &     3\farcs16 & 18.87 & 14.90 & 13.05 & $-$4.9 & 47.8 &$-$8.7 & 2.78e-016 \\
59 & \nodata &     11\farcs54 &     8\farcs42 & 18.27 & 14.63 & 13.05 & $-$4.9 & 47.8 &\nodata & \nodata \\
60 & \nodata &      7\farcs08 &     0\farcs58 & 18.32 & 14.56 & 13.02 & $-$4.9 & 47.0 &7.0 & 3.06e-016 \\
61 & \nodata &   $-$1\farcs53 &    23\farcs67 & 18.56 & 14.69 & 13.09 & $-$4.9 & 47.0 &\nodata & \nodata \\
62 & \nodata &      2\farcs20 &     6\farcs15 & 18.10 & 14.57 & 13.04 & $-$4.8 & 46.4 &$-$0.5 & 2.90e-016 \\
63 & \nodata &      7\farcs60 &  $-$2\farcs92 & 18.21 & 14.64 & 13.15 & $-$4.8 & 46.0 &$-$22.0 & 2.74e-016 \\
64 & \nodata &      2\farcs91 &     6\farcs05 & 18.25 & 14.69 & 13.13 & $-$4.8 & 45.9 &0.7 & 2.62e-016 \\
65 & \nodata &   $-$2\farcs34 &     1\farcs23 & 18.67 & 14.90 & 13.16 & $-$4.8 & 45.7 &$-$5.6 & 2.61e-016 \\
66 & \nodata &      3\farcs57 &     2\farcs34 & 18.32 & 14.63 & 13.11 & $-$4.8 & 44.9 &$-$1.4 & 2.98e-016 \\
67 & \nodata &      2\farcs82 &     7\farcs66 & 18.68 & 14.94 & 13.35 & $-$4.7 & 43.8 &$-$4.2 & 2.16e-016 \\
68 & \nodata &      7\farcs82 &     4\farcs61 & 17.74 & 14.33 & 12.93 & $-$4.7 & 43.7 &$-$5.1 & 3.53e-016 \\
69 & \nodata &      7\farcs72 &     2\farcs44 & 18.10 & 14.52 & 13.06 & $-$4.7 & 43.5 &$-$0.4 & 3.13e-016 \\
70 & \nodata &   $-$0\farcs10 &  $-$0\farcs63 & 18.89 & 14.94 & 13.30 & $-$4.7 & 43.2 &$-$9.4 & 2.34e-016 \\
71 & \nodata &   $-$2\farcs63 &     6\farcs33 & 19.28 & 15.32 & 13.62 & $-$4.7 & 43.1 &$-$3.7 & 1.72e-016 \\
72 & \nodata &      6\farcs88 &     2\farcs49 & 18.21 & 14.62 & 13.13 & $-$4.6 & 42.5 &32.1 & 2.89e-016 \\
73 & \nodata &      5\farcs50 &  $-$2\farcs82 & 18.81 & 14.96 & 13.35 & $-$4.6 & 42.3 &$-$12.4 & 2.37e-016 \\
74 & \nodata &      9\farcs63 &  $-$4\farcs73 & 18.73 & 14.98 & 13.36 & $-$4.6 & 42.2 &\nodata & \nodata \\
75 & \nodata &      7\farcs42 &    11\farcs51 & 19.83 & 15.32 & 13.36 & $-$4.6 & 42.2 &\nodata & \nodata \\
76 & \nodata &      1\farcs07 &     5\farcs70 & 18.81 & 14.98 & 13.38 & $-$4.6 & 41.9 &$-$14.2 & 2.04e-016 \\
77 & \nodata &      6\farcs71 &     2\farcs93 & 18.40 & 14.74 & 13.25 & $-$4.6 & 41.3 &$-$1.4 & 2.61e-016 \\
78 & \nodata &     22\farcs07 &  $-$1\farcs95 & 19.03 & 15.08 & 13.44 & $-$4.5 & 40.8 &\nodata & \nodata \\
79 & \nodata &   $-$2\farcs45 &     4\farcs65 & 19.91 & 15.41 & 13.46 & $-$4.5 & 40.5 &$-$9.4 & 1.98e-016 \\
80 & \nodata &      4\farcs33 &     3\farcs30 & 18.45 & 14.92 & 13.47 & $-$4.5 & 40.3 &$-$5.2 & 2.25e-016 \\
81 & \nodata &      5\farcs95 &  $-$1\farcs48 & 18.74 & 15.02 & 13.48 & $-$4.5 & 40.1 &$-$10.9 & 2.00e-016 \\
82 & \nodata &      6\farcs26 &     3\farcs63 & 18.35 & 14.93 & 13.40 & $-$4.5 & 40.1 &$-$2.2 & 2.24e-016 \\
83 & \nodata &   $-$2\farcs62 &     2\farcs09 & 19.21 & 15.34 & 13.53 & $-$4.4 & 39.3 &$-$9.4 & 1.80e-016 \\
84 & \nodata &      3\farcs50 &     6\farcs48 & 18.70 & 15.08 & 13.54 & $-$4.4 & 39.1 &$-$13.6 & 1.95e-016 \\
85 & \nodata &      2\farcs94 &     4\farcs16 & 18.50 & 15.03 & 13.54 & $-$4.4 & 39.1 &$-$8.8 & 1.85e-016 \\
86 & \nodata &      9\farcs53 & $-$10\farcs83 & 18.33 & 15.00 & 13.56 & $-$4.4 & 38.7 &\nodata & \nodata \\
87 & \nodata &      3\farcs39 &     6\farcs18 & 18.71 & 15.10 & 13.60 & $-$4.4 & 38.1 &$-$5.4 & 1.74e-016 \\
88 & \nodata &   $-$7\farcs98 &     5\farcs53 & 20.11 & 15.55 & 13.62 & $-$4.3 & 37.8 &\nodata & \nodata \\
89 & \nodata &      9\farcs14 &     5\farcs61 & 18.48 & 15.07 & 13.65 & $-$4.3 & 37.3 &$-$7.1 & 1.84e-016 \\
90 & \nodata &      4\farcs08 & $-$10\farcs21 & 19.37 & 15.38 & 13.68 & $-$4.3 & 36.9 &$-$20.6 & 1.47e-016 \\
91 & \nodata &   $-$3\farcs70 &     5\farcs24 & 19.36 & 15.52 & 13.69 & $-$4.3 & 36.8 &$-$6.8 & 1.39e-016 \\
92 & \nodata &      3\farcs39 &     1\farcs31 & 18.59 & 14.95 & 13.48 & $-$4.3 & 36.8 &$-$2.9 & 2.18e-016 \\
93 & \nodata &      1\farcs52 &  $-$0\farcs02 & 19.18 & 15.43 & 13.81 & $-$4.3 & 36.5 &0.4 & 1.57e-016 \\
94 & \nodata &      5\farcs22 &  $-$5\farcs27 & 19.81 & 15.91 & 14.14 & $-$4.3 & 36.5 &$-$1.1 & 1.06e-016 \\
95 & \nodata &      4\farcs18 &     0\farcs76 & 18.73 & 15.18 & 13.71 & $-$4.3 & 36.4 &$-$8.5 & 1.67e-016 \\
96 & \nodata &      6\farcs10 &     3\farcs03 & 18.47 & 15.01 & 13.54 & $-$4.3 & 36.3 &$-$5.3 & 2.03e-016 \\
97 & \nodata &     14\farcs78 &     6\farcs19 & 19.02 & 15.33 & 13.73 & $-$4.2 & 36.2 &\nodata & \nodata \\
98 & \nodata &     11\farcs74 & $-$14\farcs30 & 19.16 & 15.36 & 13.75 & $-$4.2 & 35.8 &\nodata & \nodata \\
99 & \nodata &      4\farcs01 &     5\farcs67 & 19.42 & 15.35 & 13.76 & $-$4.2 & 35.6 &$-$47.3 & 1.44e-016 \\
100 & \nodata &     10\farcs53 &     5\farcs15 & 18.71 & 15.21 & 13.77 & $-$4.2 & 35.6 &\nodata & \nodata \\
101 & \nodata &      8\farcs59 &     4\farcs39 & 18.69 & 15.28 & 13.78 & $-$4.2 & 35.4 &$-$10.0 & 1.63e-016 \\
102 & \nodata &      8\farcs09 &  $-$6\farcs04 & 19.19 & 15.35 & 13.78 & $-$4.2 & 35.4 &$-$9.7 & 1.51e-016 \\
103 & \nodata &   $-$7\farcs64 &     0\farcs18 & 21.30 & 16.45 & 14.53 & $-$4.2 & 35.3 &3.0 & 6.98e-017 \\
104 & \nodata &      4\farcs17 & $-$21\farcs01 & 19.13 & 15.39 & 13.79 & $-$4.2 & 35.3 &\nodata & \nodata \\
105 & \nodata &      4\farcs41 & $-$16\farcs84 & 20.07 & 15.71 & 13.81 & $-$4.2 & 34.9 &\nodata & \nodata \\
106 & \nodata &   $-$4\farcs92 &  $-$0\farcs47 & 99.00 & 15.47 & 13.82 & $-$4.1 & 34.8 &$-$169.5 & 1.41e-016 \\
107 & \nodata &   $-$5\farcs10 &  $-$8\farcs23 & 19.11 & 15.39 & 13.87 & $-$4.1 & 34.1 &\nodata & \nodata \\
108 & \nodata &      7\farcs22 &    11\farcs83 & 19.29 & 15.54 & 13.89 & $-$4.1 & 33.7 &\nodata & \nodata \\
109 & \nodata &      3\farcs42 &     2\farcs06 & 19.19 & 15.44 & 13.91 & $-$4.1 & 33.5 &$-$10.2 & 1.42e-016 \\
110 & \nodata &      4\farcs13 &     6\farcs41 & 19.04 & 15.44 & 13.93 & $-$4.0 & 33.2 &$-$11.2 & 1.25e-016 \\
111 & \nodata &      0\farcs65 &    18\farcs90 & 19.47 & 15.67 & 13.94 & $-$4.0 & 33.1 &\nodata & \nodata \\
112 & \nodata &      3\farcs25 &     6\farcs40 & 18.94 & 15.40 & 13.87 & $-$4.0 & 32.9 &$-$0.3 & 1.31e-016 \\
113 & \nodata &   $-$3\farcs19 &     5\farcs29 & 19.80 & 15.70 & 13.97 & $-$4.0 & 32.7 &$-$8.0 & 1.14e-016 \\
114 & \nodata &      7\farcs15 &     3\farcs69 & 18.97 & 15.46 & 13.99 & $-$4.0 & 32.3 &$-$6.3 & 1.27e-016 \\
115 & \nodata &      2\farcs32 &     2\farcs72 & 19.27 & 15.32 & 13.83 & $-$4.0 & 32.3 &2.9 & 1.42e-016 \\
116 & \nodata &      3\farcs64 &    16\farcs48 & 19.53 & 15.71 & 14.01 & $-$4.0 & 32.1 &\nodata & \nodata \\
117 & \nodata &      6\farcs51 &     3\farcs32 & 19.14 & 15.54 & 14.06 & $-$3.9 & 31.4 &$-$8.0 & 1.25e-016 \\
118 & \nodata &      2\farcs54 &  $-$3\farcs21 & 19.66 & 15.82 & 14.08 & $-$3.9 & 31.1 &$-$8.0 & 1.16e-016 \\
119 & \nodata &      7\farcs65 &     0\farcs01 & 19.29 & 15.62 & 14.06 & $-$3.9 & 31.1 &0.2 & 1.13e-016 \\
120 & \nodata &      5\farcs68 &  $-$7\farcs20 & 19.82 & 15.86 & 14.09 & $-$3.9 & 31.1 &$-$14.9 & 1.10e-016 \\
121 & \nodata &   $-$0\farcs73 &     3\farcs12 & 19.51 & 15.66 & 14.09 & $-$3.9 & 31.0 &$-$11.4 & 1.06e-016 \\
122 & \nodata &      1\farcs29 &     5\farcs48 & 19.24 & 15.58 & 14.09 & $-$3.9 & 30.9 &$-$13.6 & 1.17e-016 \\
123 &    B14 &      5\farcs58 &     0\farcs41 & 99.00 & 15.96 & 14.11 & $-$3.9 & 30.8 &$-$16.8 & 1.04e-016 \\
124 & \nodata &      6\farcs02 &     5\farcs50 & 19.49 & 15.66 & 14.11 & $-$3.9 & 30.7 &$-$8.5 & 8.40e-017 \\
125 & \nodata &     16\farcs14 &     1\farcs01 & 20.41 & 16.03 & 14.15 & $-$3.8 & 30.1 &\nodata & \nodata \\
126 & \nodata &      8\farcs80 &    19\farcs13 & 19.12 & 15.73 & 14.16 & $-$3.8 & 30.0 &\nodata & \nodata \\
127 & \nodata &     12\farcs14 &  $-$8\farcs76 & 20.41 & 15.98 & 14.16 & $-$3.8 & 30.0 &\nodata & \nodata \\
128 & \nodata &     12\farcs56 & $-$13\farcs77 & 19.69 & 15.86 & 14.18 & $-$3.8 & 29.8 &\nodata & \nodata \\
129 & \nodata &   $-$9\farcs61 &    10\farcs04 & 20.61 & 16.16 & 14.20 & $-$3.8 & 29.4 &\nodata & \nodata \\
130 & \nodata &      4\farcs53 &  $-$4\farcs74 & 19.94 & 15.97 & 14.21 & $-$3.8 & 29.3 &$-$12.9 & 1.01e-016 \\
131 &    B29 &   $-$1\farcs28 &  $-$4\farcs70 & 20.04 & 16.01 & 14.23 & $-$3.7 & 29.1 &$-$17.2 & 8.29e-017 \\
132 & \nodata &      7\farcs04 &    20\farcs08 & 19.61 & 15.90 & 14.27 & $-$3.7 & 28.5 &\nodata & \nodata \\
133 & \nodata &      6\farcs72 &     7\farcs79 & 19.25 & 15.90 & 14.27 & $-$3.7 & 28.5 &$-$14.2 & 1.03e-016 \\
134 & \nodata &      3\farcs14 & $-$13\farcs78 & 20.15 & 16.18 & 14.28 & $-$3.7 & 28.4 &\nodata & \nodata \\
135 & \nodata &      1\farcs35 &  $-$2\farcs97 & 19.91 & 16.06 & 14.29 & $-$3.7 & 28.3 &$-$13.2 & 9.86e-017 \\
136 & \nodata &   $-$2\farcs41 &  $-$5\farcs08 & 20.53 & 16.21 & 14.30 & $-$3.7 & 28.1 &$-$6.4 & 8.96e-017 \\
137 & \nodata &     16\farcs11 &     1\farcs50 & 20.92 & 16.32 & 14.32 & $-$3.7 & 27.9 &\nodata & \nodata \\
138 & \nodata &   $-$0\farcs63 & $-$16\farcs45 & 20.61 & 16.37 & 14.35 & $-$3.6 & 27.5 &\nodata & \nodata \\
139 & \nodata &     12\farcs45 &  $-$0\farcs65 & 19.75 & 15.96 & 14.36 & $-$3.6 & 27.3 &$-$10.7 & 8.77e-017 \\
140 & \nodata &      3\farcs99 &     4\farcs28 & 19.19 & 15.83 & 14.38 & $-$3.6 & 27.2 &$-$9.9 & 7.72e-017 \\
141 & \nodata &  $-$18\farcs83 &     3\farcs95 & 20.93 & 16.41 & 14.39 & $-$3.6 & 26.9 &\nodata & \nodata \\
142 & \nodata &      5\farcs64 &  $-$2\farcs17 & 19.76 & 15.97 & 14.41 & $-$3.6 & 26.7 &\nodata & \nodata \\
143 & \nodata &      5\farcs99 &     5\farcs52 & 19.69 & 16.02 & 14.43 & $-$3.5 & 26.5 &$-$8.5 & 8.40e-017 \\
144 & \nodata &     16\farcs64 &     2\farcs18 & 20.22 & 16.19 & 14.43 & $-$3.5 & 26.5 &\nodata & \nodata \\
145 & \nodata &     15\farcs57 &     2\farcs69 & 19.65 & 16.05 & 14.44 & $-$3.5 & 26.3 &\nodata & \nodata \\
146 & \nodata &   $-$5\farcs16 &     7\farcs86 & 19.96 & 16.07 & 14.45 & $-$3.5 & 26.2 &$-$9.7 & 7.53e-017 \\
147 & \nodata &     14\farcs38 &     2\farcs76 & 19.63 & 16.11 & 14.46 & $-$3.5 & 26.2 &\nodata & \nodata \\
148 & \nodata &     20\farcs04 &  $-$5\farcs66 & 19.69 & 16.07 & 14.47 & $-$3.5 & 26.0 &\nodata & \nodata \\
149 & \nodata &      5\farcs54 &    21\farcs20 & 20.39 & 16.16 & 14.47 & $-$3.5 & 26.0 &\nodata & \nodata \\
150 & \nodata &      2\farcs57 &     3\farcs12 & 19.57 & 15.81 & 14.31 & $-$3.5 & 25.9 &12.8 & 8.94e-017 \\
151 & \nodata &      7\farcs48 &  $-$4\farcs87 & 19.91 & 16.09 & 14.48 & $-$3.5 & 25.9 &\nodata & \nodata \\
152 & \nodata &      5\farcs71 & $-$13\farcs80 & 20.25 & 16.28 & 14.48 & $-$3.5 & 25.8 &\nodata & \nodata \\
153 & \nodata &      1\farcs31 &     2\farcs76 & 20.15 & 16.12 & 14.51 & $-$3.5 & 25.5 &$-$5.7 & 7.45e-017 \\
154 & \nodata &   $-$3\farcs52 & $-$10\farcs44 & 20.43 & 16.34 & 14.52 & $-$3.5 & 25.4 &\nodata & \nodata \\
155 & \nodata &      2\farcs13 &     3\farcs63 & 19.74 & 16.00 & 14.52 & $-$3.4 & 25.4 &$-$12.2 & 7.38e-017 \\
156 & \nodata &      5\farcs02 &    20\farcs61 & 20.38 & 16.30 & 14.54 & $-$3.4 & 25.2 &\nodata & \nodata \\
157 & \nodata &   $-$0\farcs53 &  $-$1\farcs42 & 20.12 & 16.33 & 14.55 & $-$3.4 & 25.0 &$-$15.2 & 7.83e-017 \\
158 & \nodata &     10\farcs98 &     0\farcs77 & 99.00 & 16.33 & 14.56 & $-$3.4 & 24.9 &$-$5.2 & 6.58e-017 \\
159 & \nodata &      8\farcs17 &     7\farcs07 & 20.43 & 16.32 & 14.57 & $-$3.4 & 24.8 &\nodata & \nodata \\
160 & \nodata &      3\farcs16 &     6\farcs95 & 19.81 & 16.11 & 14.59 & $-$3.4 & 24.6 &$-$12.4 & 7.01e-017 \\
161 & \nodata &      8\farcs35 & $-$17\farcs92 & 20.86 & 16.48 & 14.59 & $-$3.4 & 24.6 &\nodata & \nodata \\
162 & \nodata &      7\farcs36 & $-$16\farcs34 & 20.12 & 16.27 & 14.59 & $-$3.4 & 24.6 &\nodata & \nodata \\
163 & \nodata &      2\farcs31 &  $-$0\farcs64 & 20.01 & 16.17 & 14.60 & $-$3.4 & 24.5 &4.2 & 3.17e-017 \\
164 & \nodata &      7\farcs91 &    14\farcs29 & 19.75 & 16.15 & 14.61 & $-$3.4 & 24.4 &\nodata & \nodata \\
165 & \nodata &     11\farcs83 &     7\farcs40 & 19.62 & 16.14 & 14.65 & $-$3.3 & 23.9 &\nodata & \nodata \\
166 & \nodata &      5\farcs46 &     3\farcs19 & 19.80 & 16.20 & 14.69 & $-$3.3 & 23.5 &$-$10.4 & 7.22e-017 \\
167 & \nodata &      6\farcs22 &     5\farcs03 & 19.34 & 15.86 & 14.42 & $-$3.3 & 23.4 &9.8 & 8.94e-017 \\
168 & \nodata &      3\farcs09 &  $-$7\farcs45 & 20.48 & 16.55 & 14.70 & $-$3.3 & 23.4 &$-$13.4 & 6.18e-017 \\
169 & \nodata &      1\farcs54 &     2\farcs50 & 20.09 & 16.32 & 14.72 & $-$3.3 & 23.3 &$-$6.3 & 6.15e-017 \\
170 &   AR11  &      5\farcs83 &     1\farcs80 & 19.87 & 16.16 & 14.74 & $-$3.2 & 23.1 &$-$12.2 & 6.59e-017 \\
171 & \nodata &  $-$12\farcs94 &    10\farcs63 & 20.39 & 16.48 & 14.76 & $-$3.2 & 22.8 &\nodata & \nodata \\
172 & \nodata &   $-$0\farcs37 &     5\farcs24 & 20.22 & 16.31 & 14.76 & $-$3.2 & 22.8 &$-$4.7 & 5.87e-017 \\
173 & \nodata &   $-$0\farcs74 &     2\farcs52 & 20.13 & 16.30 & 14.80 & $-$3.2 & 22.5 &$-$15.9 & 5.96e-017 \\
174 & \nodata &      6\farcs38 &  $-$5\farcs44 & 20.30 & 16.53 & 14.81 & $-$3.2 & 22.4 &$-$9.2 & 5.83e-017 \\
175 & \nodata &      2\farcs67 &    19\farcs47 & 20.65 & 16.56 & 14.82 & $-$3.1 & 22.2 &\nodata & \nodata \\
176 & \nodata &      0\farcs30 &  $-$1\farcs09 & 20.34 & 16.66 & 14.83 & $-$3.1 & 22.1 &$-$13.4 & 5.83e-017 \\
177 & \nodata &      4\farcs52 &     5\farcs89 & 20.07 & 16.31 & 14.83 & $-$3.1 & 22.1 &$-$7.8 & 5.61e-017 \\
178 & \nodata &      7\farcs41 &    15\farcs81 & 21.14 & 16.69 & 14.85 & $-$3.1 & 22.0 &\nodata & \nodata \\
179 & \nodata &     14\farcs40 &     1\farcs39 & 20.64 & 16.53 & 14.85 & $-$3.1 & 21.9 &\nodata & \nodata \\
180 & \nodata &      0\farcs31 &    23\farcs12 & 20.77 & 16.65 & 14.89 & $-$3.1 & 21.6 &\nodata & \nodata \\
181 & \nodata &      4\farcs04 &    15\farcs06 & 20.08 & 16.43 & 14.90 & $-$3.1 & 21.4 &\nodata & \nodata \\
182 & \nodata &      4\farcs29 &  $-$0\farcs15 & 19.92 & 16.17 & 14.70 & $-$3.1 & 21.3 &7.8 & 6.92e-017 \\
183 & \nodata &   $-$5\farcs10 & $-$19\farcs55 & 21.11 & 16.74 & 14.93 & $-$3.0 & 21.2 &\nodata & \nodata \\
184 & \nodata &   $-$3\farcs31 &  $-$6\farcs67 & 21.23 & 16.90 & 14.94 & $-$3.0 & 21.0 &$-$11.0 & 4.90e-017 \\
185 & \nodata &      0\farcs29 &  $-$4\farcs80 & 20.71 & 16.65 & 14.95 & $-$3.0 & 20.9 &$-$10.4 & 5.32e-017 \\
186 & \nodata &   $-$3\farcs40 &  $-$5\farcs48 & 21.36 & 16.91 & 14.95 & $-$3.0 & 20.9 &$-$18.3 & 4.43e-017 \\
187 & \nodata &      3\farcs71 &  $-$0\farcs54 & 20.14 & 16.42 & 14.96 & $-$3.0 & 20.8 &6.6 & 5.01e-017 \\
188 & \nodata &      8\farcs89 &     0\farcs96 & 20.15 & 16.49 & 14.97 & $-$3.0 & 20.7 &$-$9.4 & 4.97e-017 \\
189 & \nodata &      4\farcs40 &  $-$0\farcs76 & 19.90 & 16.24 & 14.77 & $-$3.0 & 20.7 &4.0 & 5.87e-017 \\
190 & \nodata &   $-$6\farcs58 &     3\farcs00 & 20.61 & 16.72 & 14.99 & $-$3.0 & 20.6 &$-$18.3 & 4.81e-017 \\
191 & \nodata &     11\farcs56 &     4\farcs59 & 20.03 & 16.51 & 15.00 & $-$3.0 & 20.5 &\nodata & \nodata \\
192 & \nodata &   $-$2\farcs78 &    17\farcs82 & 20.57 & 16.66 & 15.00 & $-$3.0 & 20.4 &\nodata & \nodata \\
193 & \nodata &      1\farcs61 &    13\farcs20 & 20.32 & 16.60 & 15.00 & $-$3.0 & 20.4 &$-$4.7 & 4.17e-017 \\
194 & \nodata &      3\farcs56 &    14\farcs83 & 21.41 & 16.96 & 15.01 & $-$3.0 & 20.3 &\nodata & \nodata \\
195 & \nodata &      3\farcs55 &     7\farcs46 & 20.13 & 16.48 & 15.03 & $-$2.9 & 20.1 &$-$8.0 & 4.44e-017 \\
196 & \nodata &      2\farcs67 &    15\farcs02 & 20.54 & 16.69 & 15.03 & $-$2.9 & 20.1 &\nodata & \nodata \\
\enddata
\tablenotetext{a
}{ID numbers were assigned by sorting the data in order of increasing M$_{\rm K}$ (decreasing luminosity), and rejecting stars outside the range 1.4 $<$ m$_{160}$ - m$_{205}$ $<$ 2.1}
\tablenotetext{b
}{Designations are taken from the following, in order of preference: (1) Nagata et al.\ (1995) (N\#), (2) Cotera et al.\ (1996) (C\#), (3) Lang et al.\ (2001) (AR\#), and (4) Blum et al.\ (2001) (B\#). Radio sources AR9-17 are newly identified in this paper, and their coordinates have been extracted from Figure 2 of Lang et al.\ (2001).}
\tablenotetext{c
}{Positions are with respect to RA(J2000) 17$^h$ 45$^m$ 50.26$^s$ DEC(J2000) $-$28$\arcdeg$ 49$\arcmin$ 22$\farcs$76 and have a relative error of $\pm$0.008\arcsec.}
\tablenotetext{d
}{M$_{\rm K}$ assumes (m$_{205}$-K)$_0$=0, d=8000~pc  \citep{rei93}, (m$_{160}$ - m$_{205}$)$_0$=$-$0.05 and A$_{\rm K}$~=~1.95 $\times$ E(\H~$-$~\K).}
\tablenotetext{e
}{\Minit\ assumes the relation between mass and magnitude for $\tau$=2.5~\Myr\ from the Geneva models with solar metallicity and enhanced mass-loss rates.}
\tablenotetext{f
}{Equivalent is calculated as $EW=[(\Fnp-\Fnc)/\Fnc]*\Delta\lambda_{\rm F187N}$, where $\Fnp$ is corrected for extinction and the intrinsic shape of the spectral energy distribution.}
\end{deluxetable}

\begin{deluxetable}{rrrrr}
\tablewidth{0pt}
\tablecaption{Model Parameters}
\tablehead{
\colhead{star\#} & \colhead{\#8}  &\colhead{\#10$_a$} & \colhead{\#10$_b$} & \colhead{\#10$_c$} \\
}  
\startdata
{log L$_*$/\Lsun}            &  6.26   &  6.27     &  6.15     & 6.24    \\
{R$_*$/\Rsun}                &  43.5   &  46.8     &  46.8     & 46.8    \\
{T$_*$~kK}                   &  32.2   &  31.1     &  29.1     & 30.7    \\
{R$_{\tau=2/3}$/\Rsun}       &  47.5   &  48.2     &  48.6     & 48.2    \\
{T$_{\rm eff}$~kK}           &  30.9   &  30.7     &  28.5     & 30.3    \\
{\Vinf \kms}                 &  1100   &  1000:    &  1000:    & 1600:   \\
{log \Mdot/\Msunyr}          & $-$4.35 &  $-$5.37  &  $-$4.79  & $-$5.21 \\
{$\beta$}                    &  1.25   &  2.25     &  1.50     & 2.25   \\
{f}                          &  0.1    &  0.1      &  1.0      & .1     \\
{m$_{\rm F205W}$}            & 10.71   &  11.44    &  11.45    & 11.43   \\
{m$_{\rm F160W}$}            & 12.50   &  13.33    &  13.35    & 13.32   \\
{m$_{\rm F110W}$}            &  16.27  &  17.41    &  17.43    & 17.40   \\
{EW$_{\rm P\alpha}$ $\AA$}   &  242.   &  54.0     &  63.3     & 54.9    \\
{A$_{\rm k}$}                & 3.34    &  3.80     &  3.75     & 3.75    \\
{H\tablenotemark{a}}         &  0.27   &  0.42     &  0.42     & 0.42    \\
{He\tablenotemark{a}}        & 0.71    &  0.56     &  0.56     & 0.56    \\
{C\tablenotemark{a}}         &  0.0002:&  0.0008:  &  0.0008:  & 0.0008: \\
{N\tablenotemark{a}}         & 0.016   &  0.006:   &  0.006:   & 0.006:  \\
{log Q(H$^+$)}               &  49.9   &  49.8     &  49.5     & 49.7    \\
{log Q(He$^+$)}              &  48.5   &  48.4     &  47.6     & 48.2    \\
\enddata
\tablenotetext{a}{Mass fraction}
\tablecomments{Quantities followed by a colon are upper limits.}
\end{deluxetable}

\newpage

\begin{figure}
\figurenum{1a}
\epsscale{1.4}
\hspace{3.75in}
\plotone{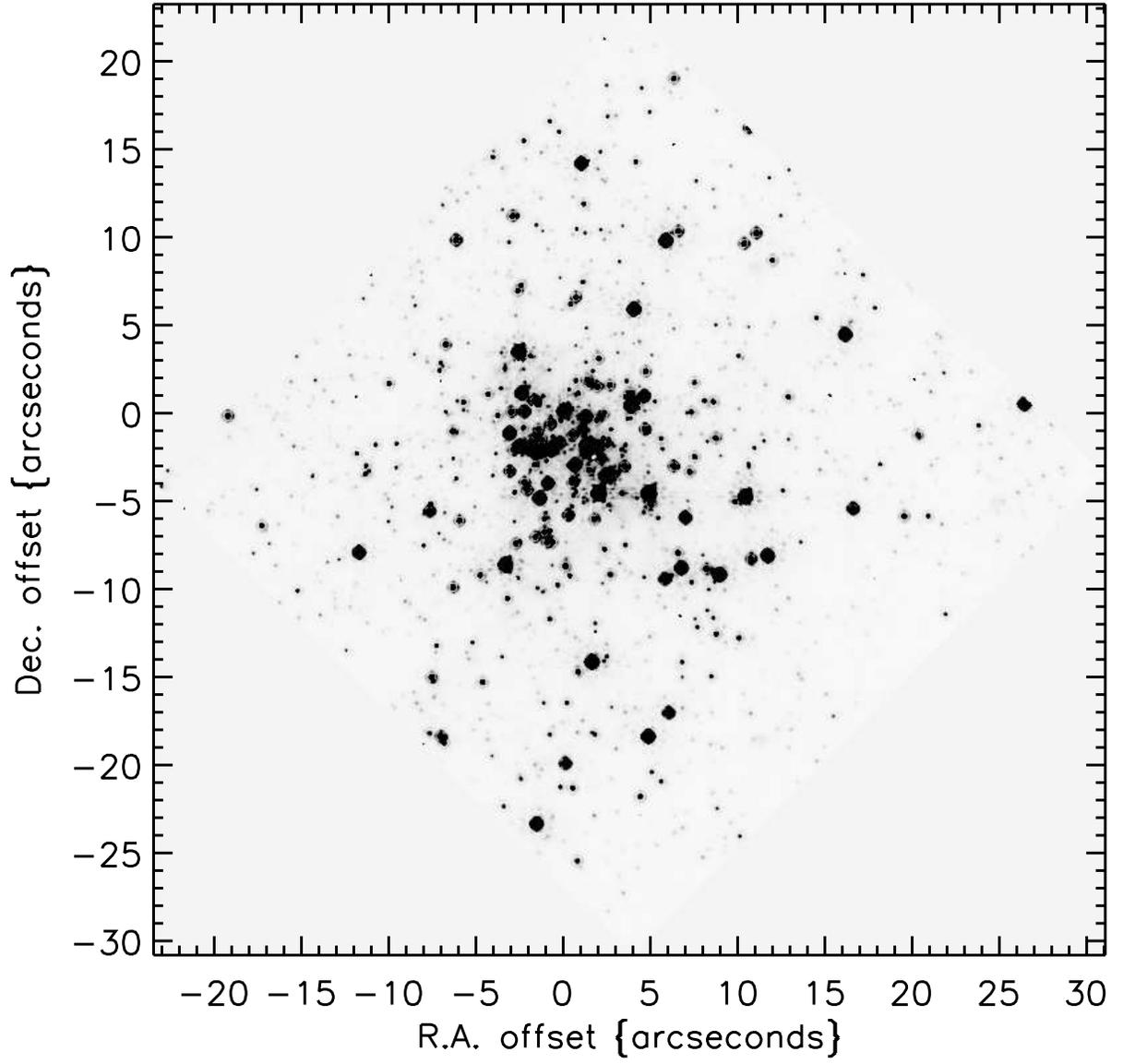}
\hspace*{4.5in} 
\vskip .2in
\caption{{\it (a)} F205W image, after processing by calnica and calnicb.}
\end{figure}

\clearpage
\begin{figure}
\figurenum{1b}
\epsscale{1.4}
\hspace{3.75in}
\plotone{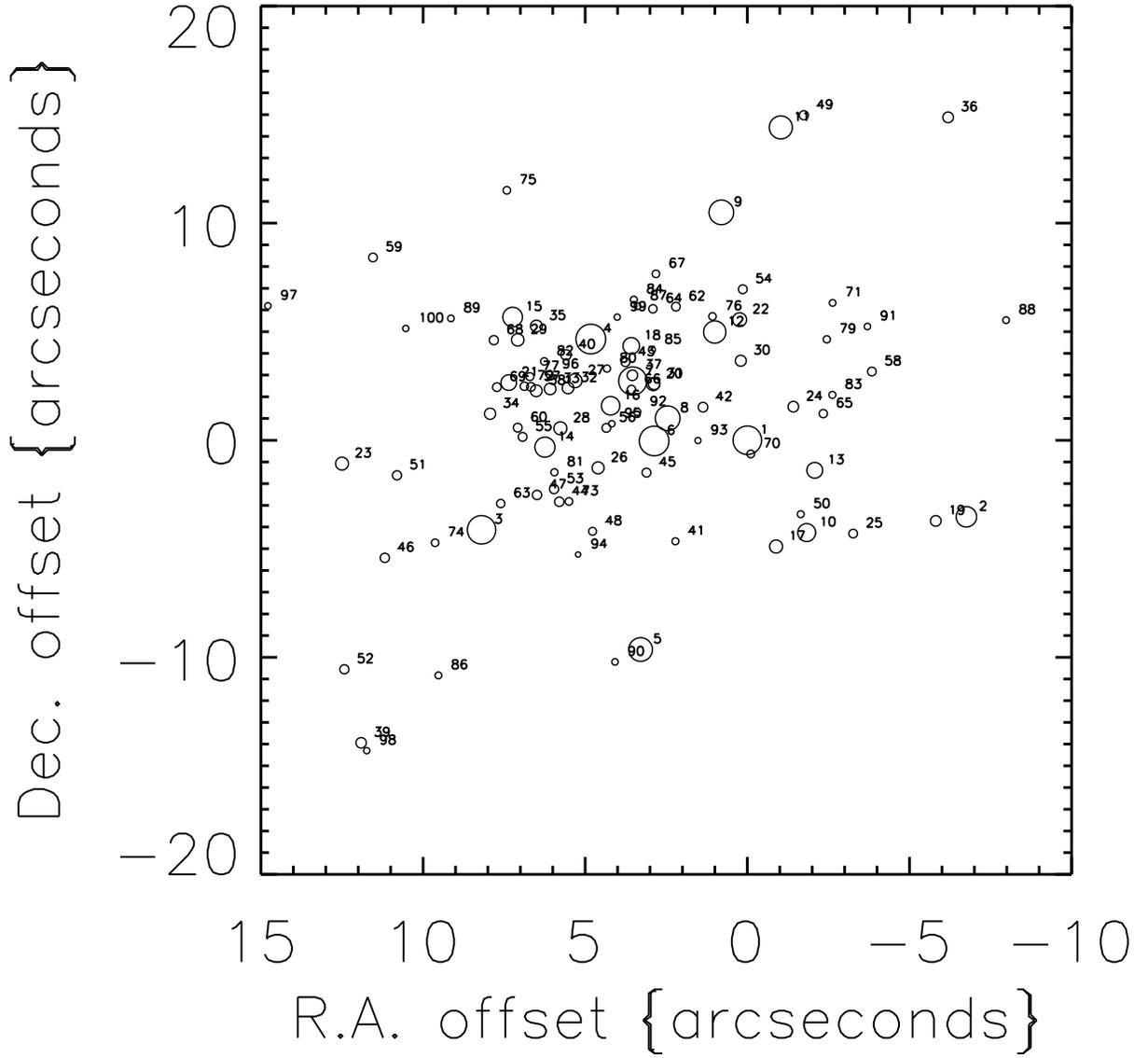}
\hspace*{4.5in} 
\vskip .2in
\caption{{\it (b)} ID's of bright stars.}
\end{figure}

\begin{figure}
\figurenum{2}
\epsscale{1.3}
\hspace{3.75in}
\plotone{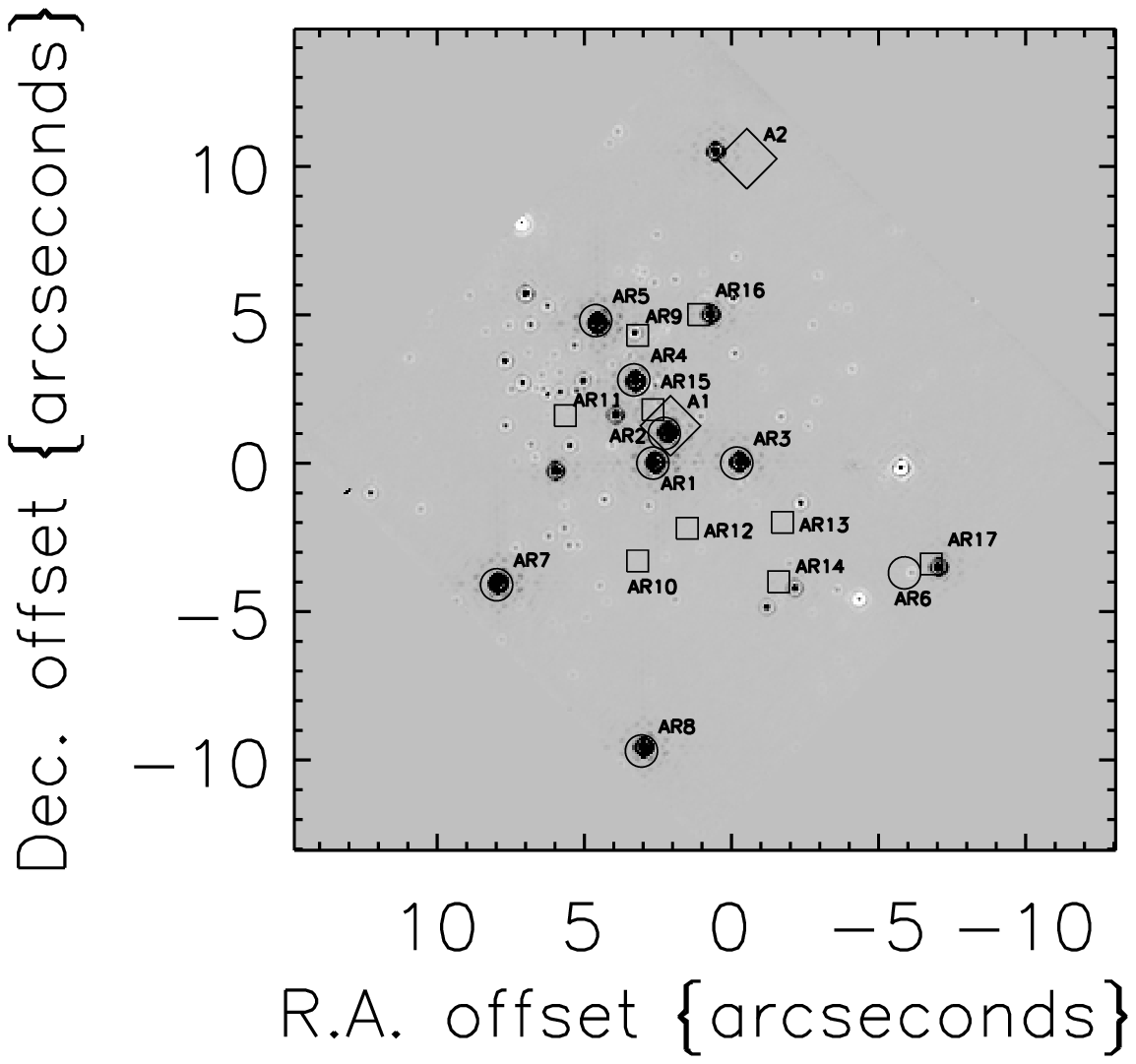}
\hspace*{4.5in} 
\vskip .2in
\caption{Difference image made by subtracting F190N image from F187N image. Positions of radio
sources AR1$-$8 ({\it circles}) are from \citet{lan01a}. Radio sources AR9$-$17 ({\it squares}) are newly 
identified in this paper, and their coordinates are taken from Figure~2 in \citet{lan01a}. Positions
of x-ray sources ({\it diamonds}) are from \citet{zad02}.}
\end{figure}

\begin{figure}
\figurenum{3}
\epsscale{1}
\hspace{3.75in}
\plottwo{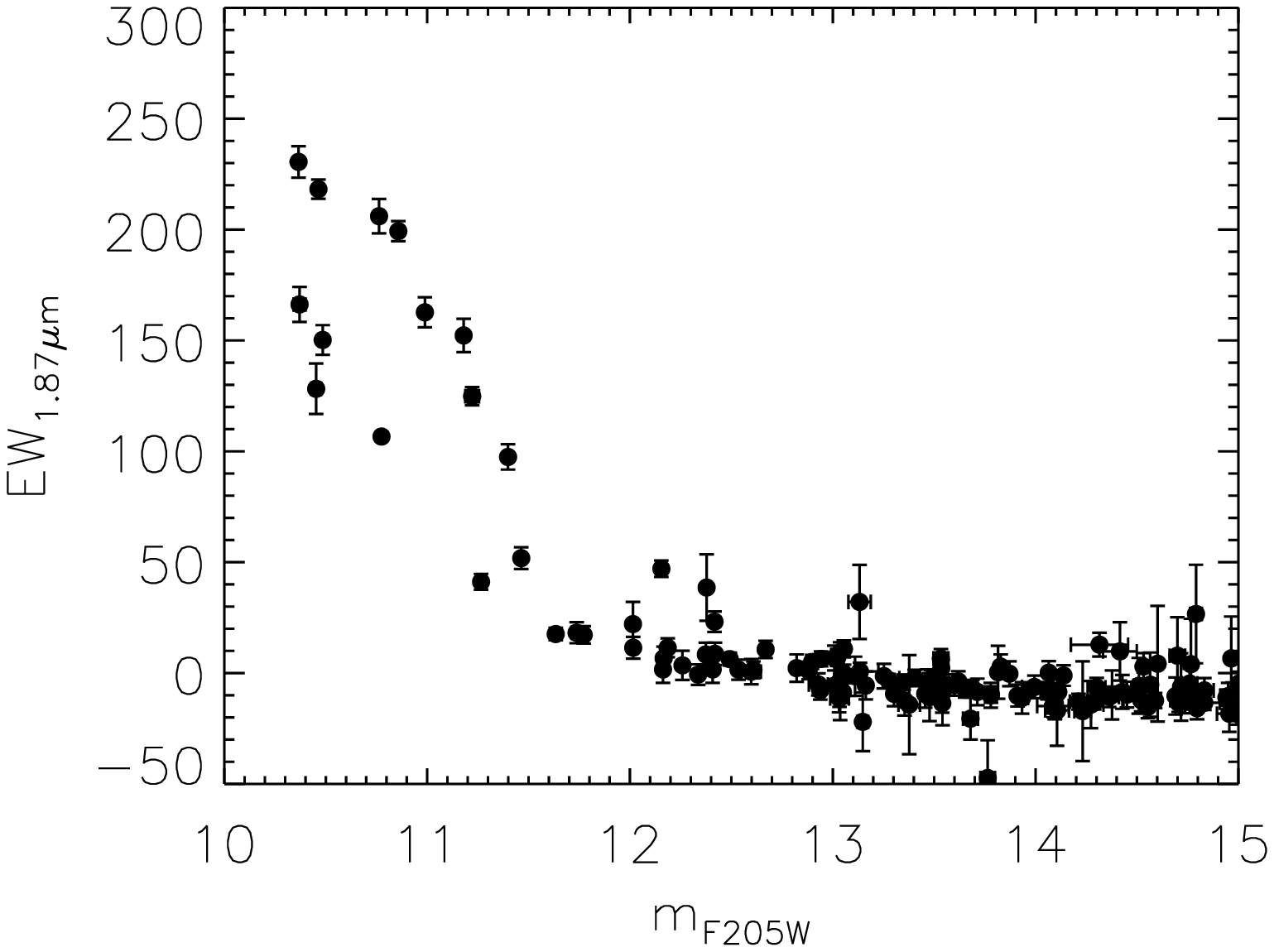}{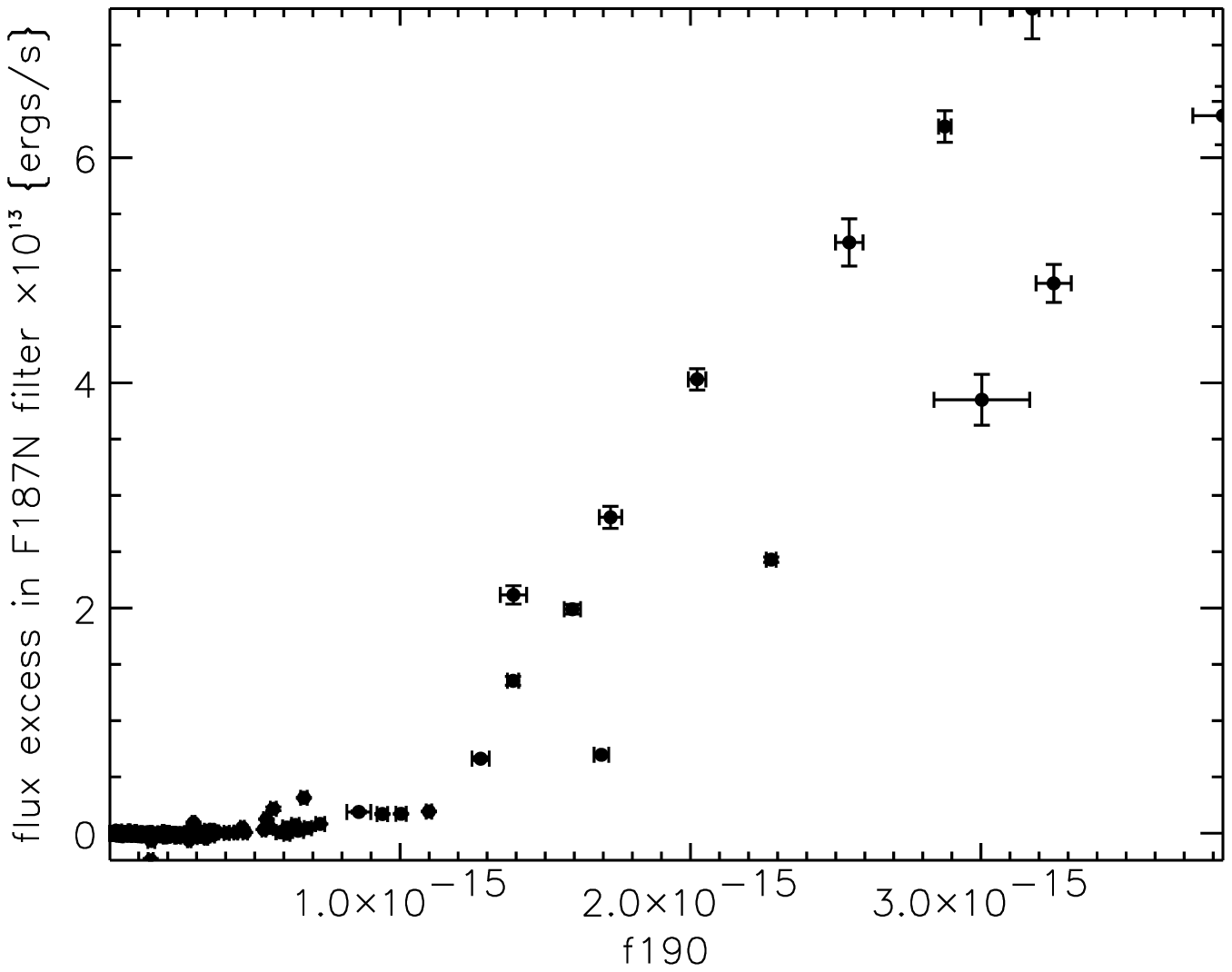}
\hspace*{4.5in} 
\vskip .2in
\caption{{\it (a)} Plot of \ew\ as a function of \mnk. {\it (b)} Linear plot of excess flux in F187N versus \Fnc.}
\end{figure}

\begin{figure}
\figurenum{4}
\centering
\includegraphics[width=10cm]{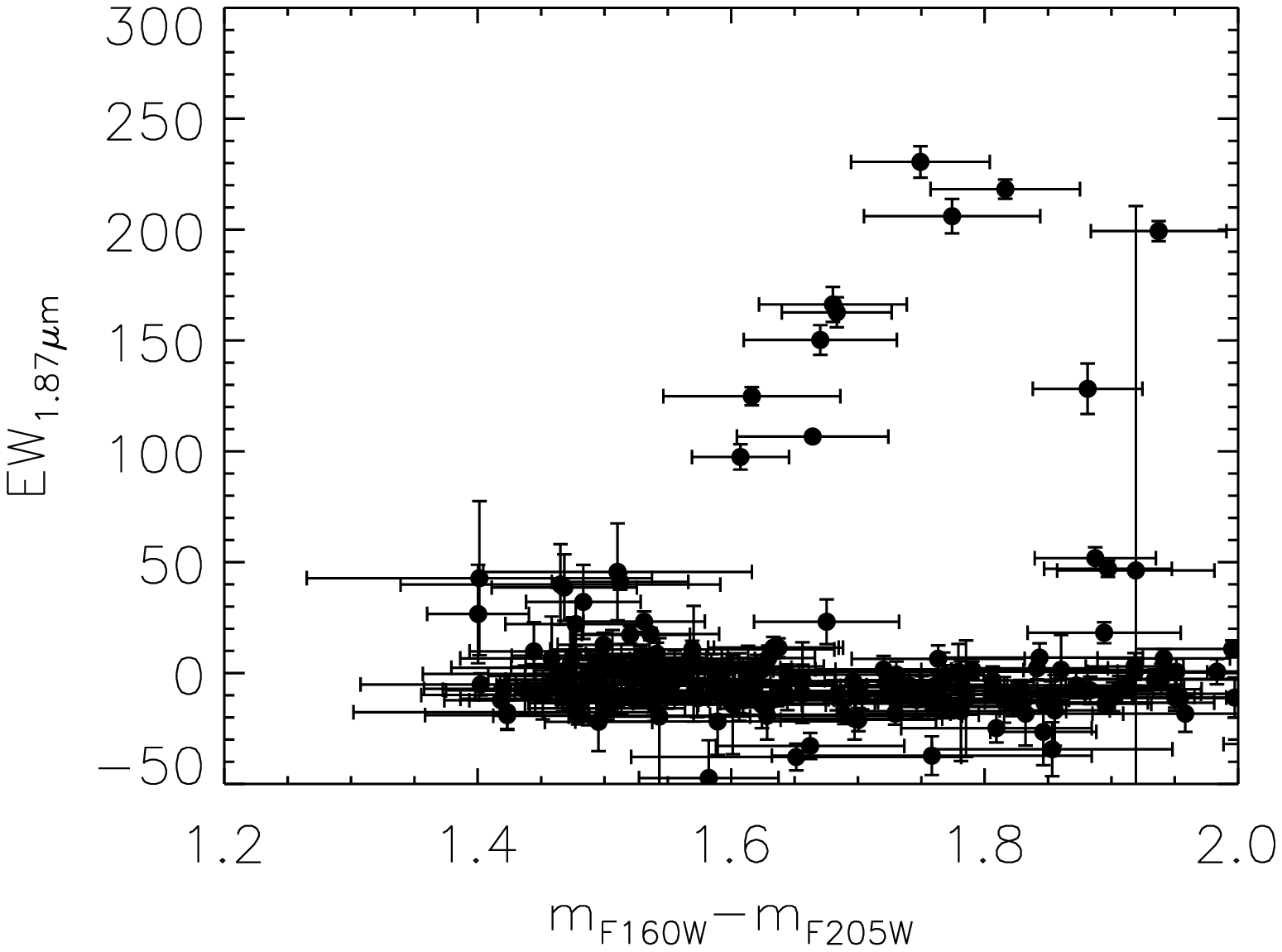}
\caption{Plot of \ew\ as a function of color in \mnh$-$\mnk. The plot includes data
points where the color error is less than 0.2.}
\end{figure}

\newpage

\begin{figure}
\figurenum{5a}
\epsscale{1.05}
\caption{Spectra of selected stars from Table~3. The spectra have not been
dereddened, and the flux scale is arbitrary. The wavelength gaps in the spectra
are due to incomplete coverage of the cross-dispersed echelle format by the
detector. The two sharp absorption features near 2.32~\micron\ are due to imperfect correction
for telluric absorption. Other sharp features (a few pixels wide), especially in the left-most
order, are similarly due to this imperfect correction, or are due to detector artifacts, c.f.\ the
feature near 2.294~\micron\ in the spectrum of star \#2.}
\end{figure}

\begin{figure}
\figurenum{5b}
\epsscale{1.05}
\caption{Same as Figure~5a.}
\end{figure}

\begin{figure}
\figurenum{5c}
\epsscale{1.05}
\hspace{3.75in}
\hspace*{4.5in} 
\vskip .2in
\caption{Same as Figure~5a. The sharp features near 2.294~\micron\ in the spectra of star
\#10 and \#14 are due to detector artifacts.}
\end{figure}

\begin{figure}
\figurenum{5d}
\epsscale{1.05}
\hspace{3.75in}
\hspace*{4.5in} 
\vskip .2in
\caption{Same as Figure~5a.}
\end{figure}

\newpage

\begin{figure}
\figurenum{5e}
\epsscale{1.05}
\hspace{3.75in}
\hspace*{4.5in} 
\vskip .2in
\caption{Same as Figure~5a.}
\end{figure}

\begin{figure}
\figurenum{5f}
\epsscale{1.05}
\hspace{3.75in}
\hspace*{4.5in} 
\vskip .2in
\caption{Same as Figure~5a.}
\end{figure}

\begin{figure}
\figurenum{5g}
\epsscale{1.05}
\hspace{3.75in}
\hspace*{4.5in} 
\vskip .2in
\caption{Same as Figure~5a.}
\end{figure}

\begin{figure}
\figurenum{5h}
\epsscale{1.05}
\hspace{3.75in}
\hspace*{4.5in} 
\vskip .2in
\caption{Same as Figure~5a.}
\end{figure}

\begin{figure}
\figurenum{5i}
\epsscale{1.05}
\hspace{3.75in}
\hspace*{4.5in} 
\vskip .2in
\caption{Same as Figure~5a.}
\end{figure}

\begin{figure}
\figurenum{6}
\epsscale{1.0}
\plotone{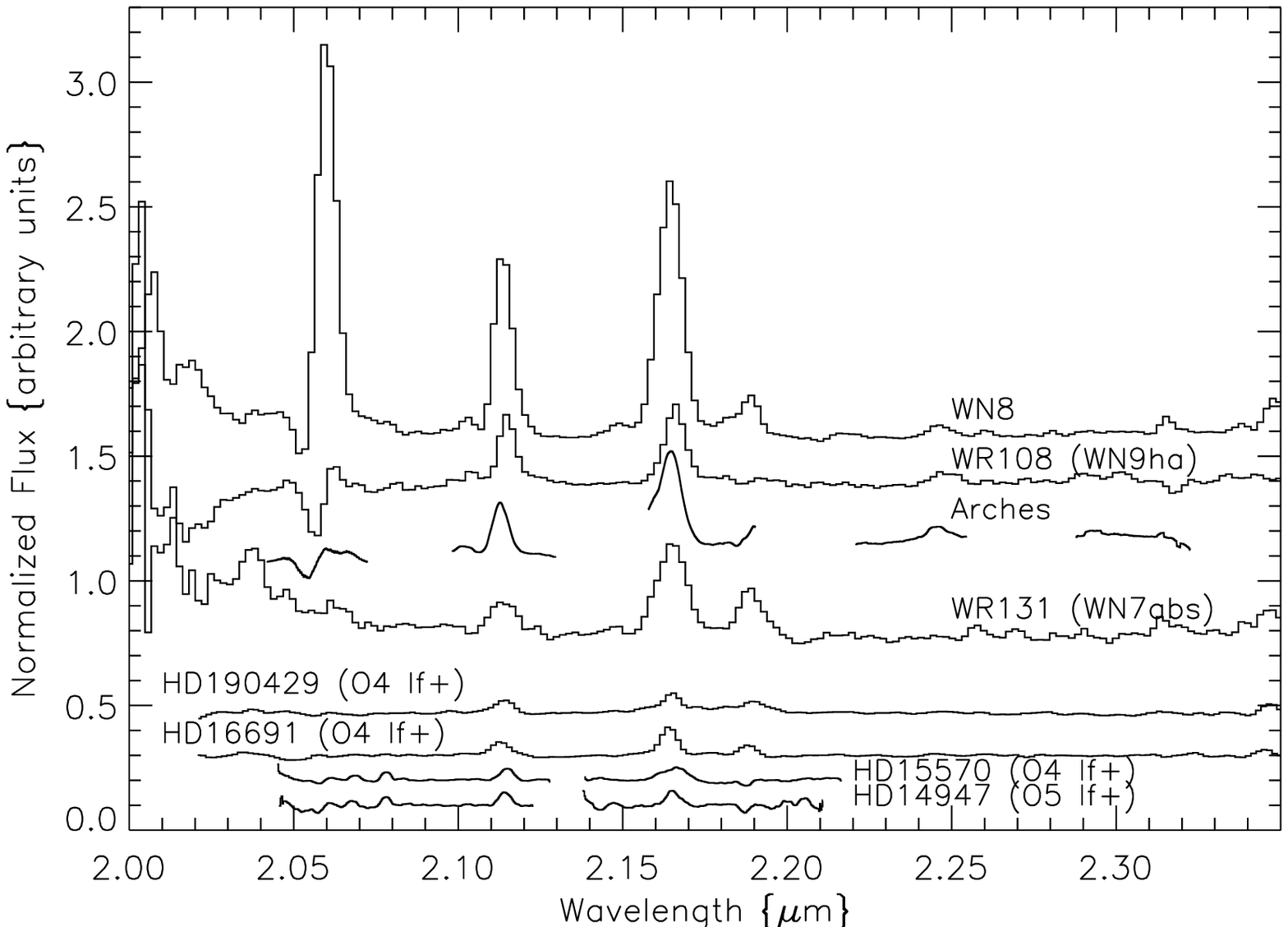}
\caption{Comparison between average spectrum of Arches emission-line stars,  
spectra of representative WN7, WN8, and WN9 stars \citep{fig97}, and spectra of individual O~If$^+$ stars 
\citep{han96}. All spectra have been smoothed to match the resolution of the
WR star spectra (R$\sim$525). Notice that the WNL and Arches stars have 
\ion{N}{3} features in common at 2.104~\micron\ and 2.25~\micron.
In addition, the equivalent-widths and relative ratios of equivalent-widths
are similar in the WNL and Arches spectra. The spectra are all on the same scale,
but shifted by a constant for presentation purposes.}
\end{figure}

\begin{figure}
\figurenum{7a}
\epsscale{1.1}
\plotone{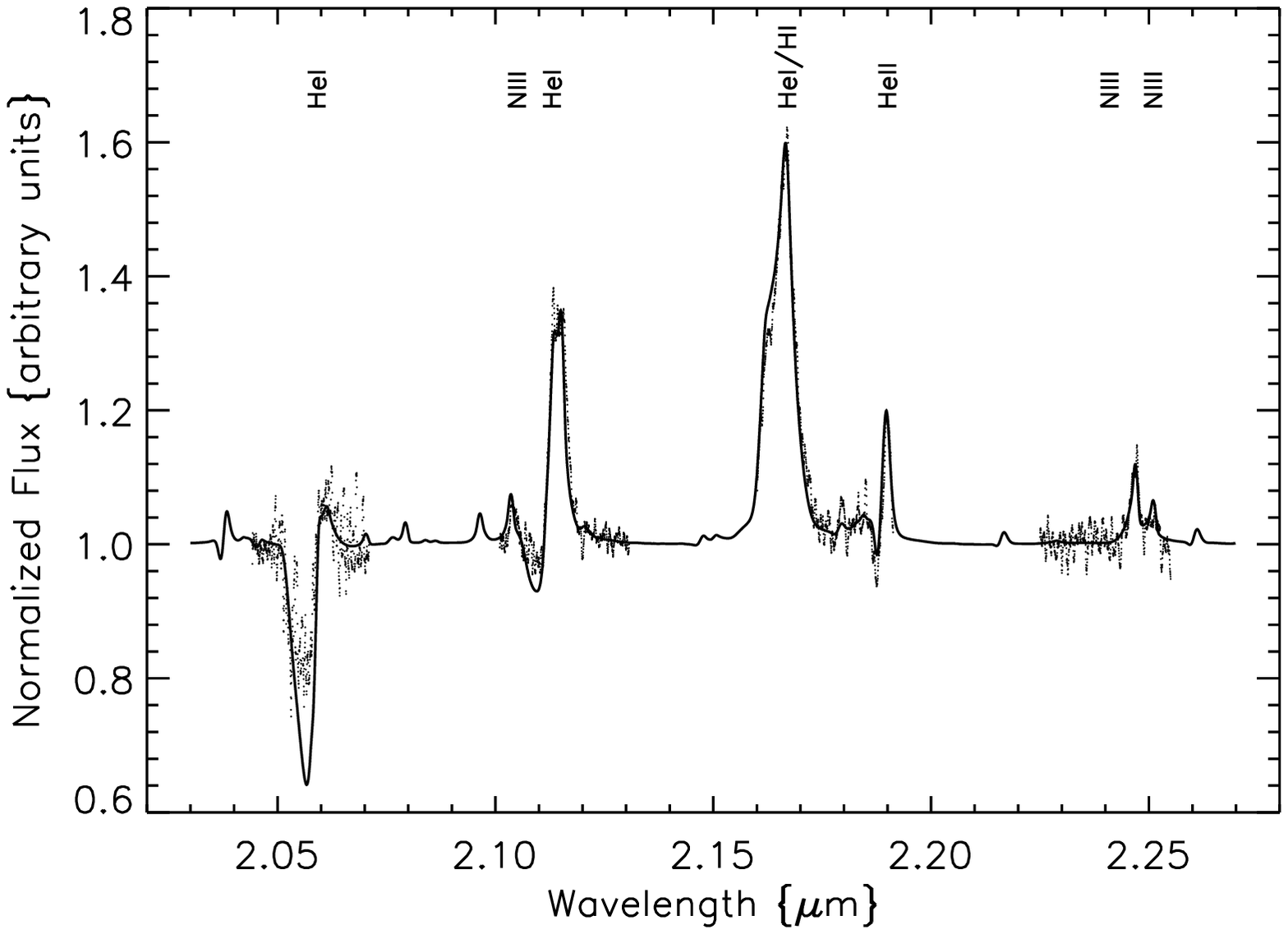}
\caption{Observed ({\it dots}) and modelled ({\it line}) spectrum of star \#8.}
\end{figure}

\begin{figure}
\figurenum{7b}
\epsscale{1.1}
\plotone{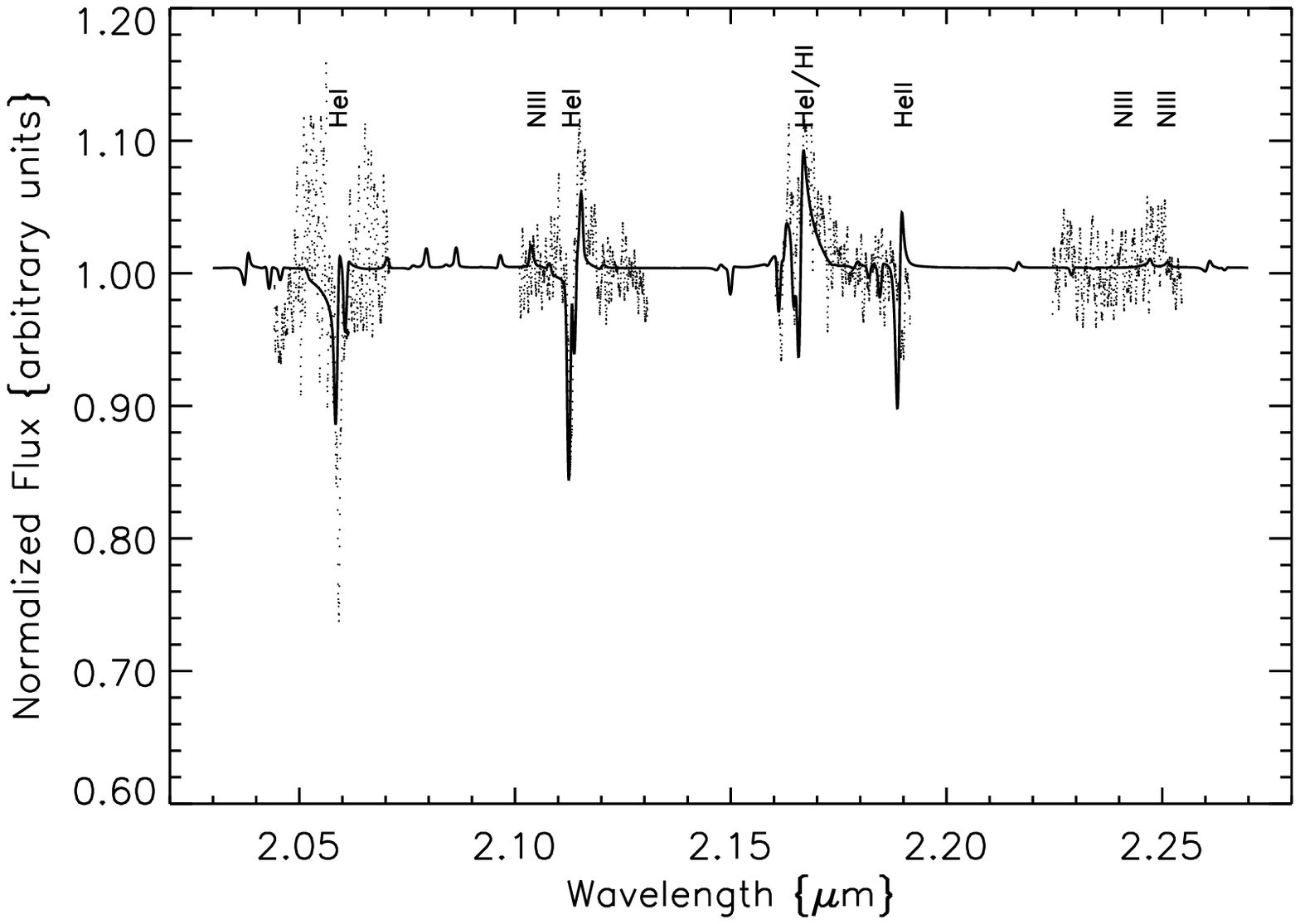}
\caption{Observed ({\it dots}) and modelled ({\it line}) spectrum of star \#10.}
\end{figure}

\clearpage 

\begin{figure}
\figurenum{8a}
\centering
\includegraphics[width=10cm]{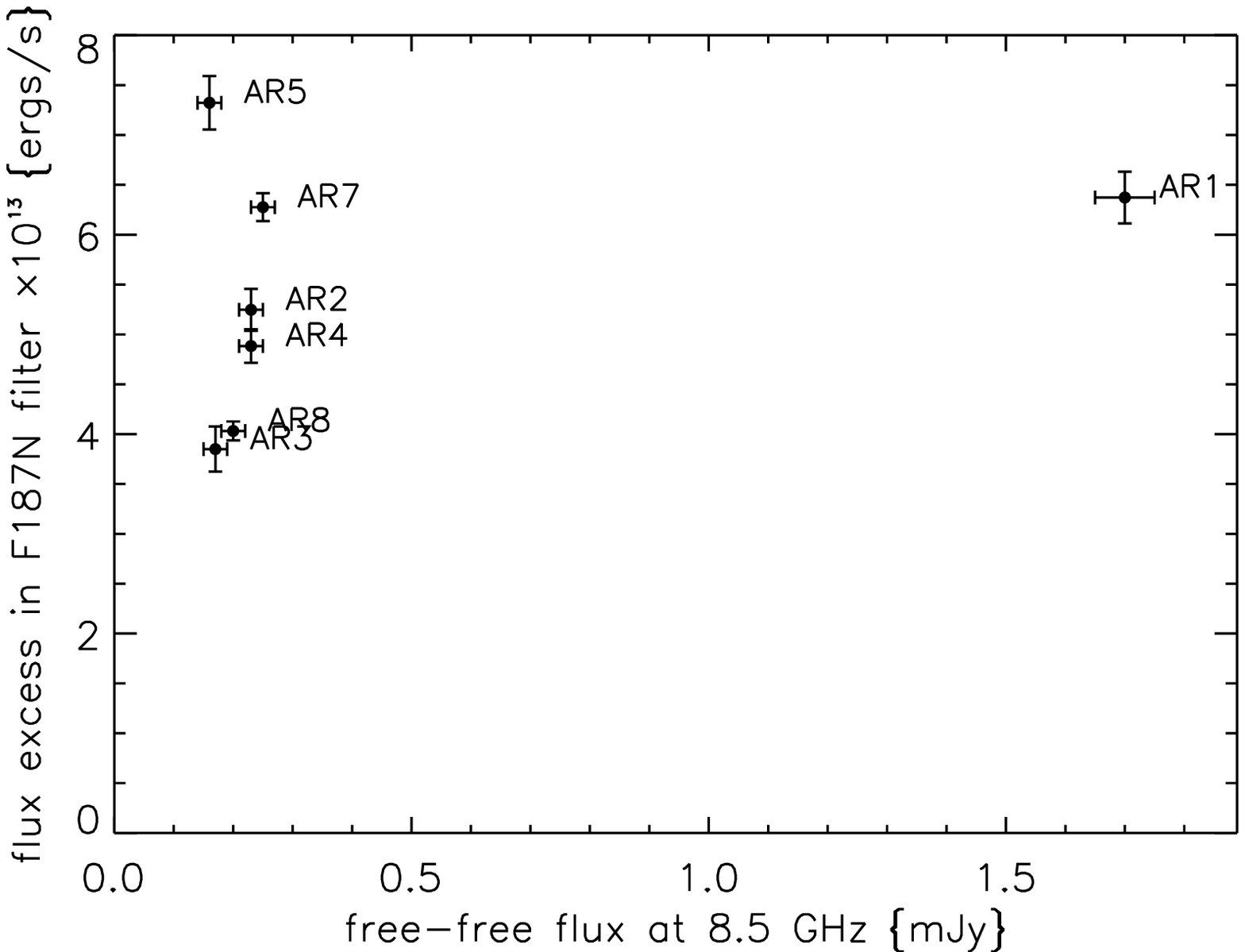}
\caption{Plot of (\Fnp$-\Fnc)\times\Delta\lambda$ versus \Frh.}
\end{figure}

\begin{figure}[p]
\figurenum{8b}
\centering
\includegraphics[width=10cm]{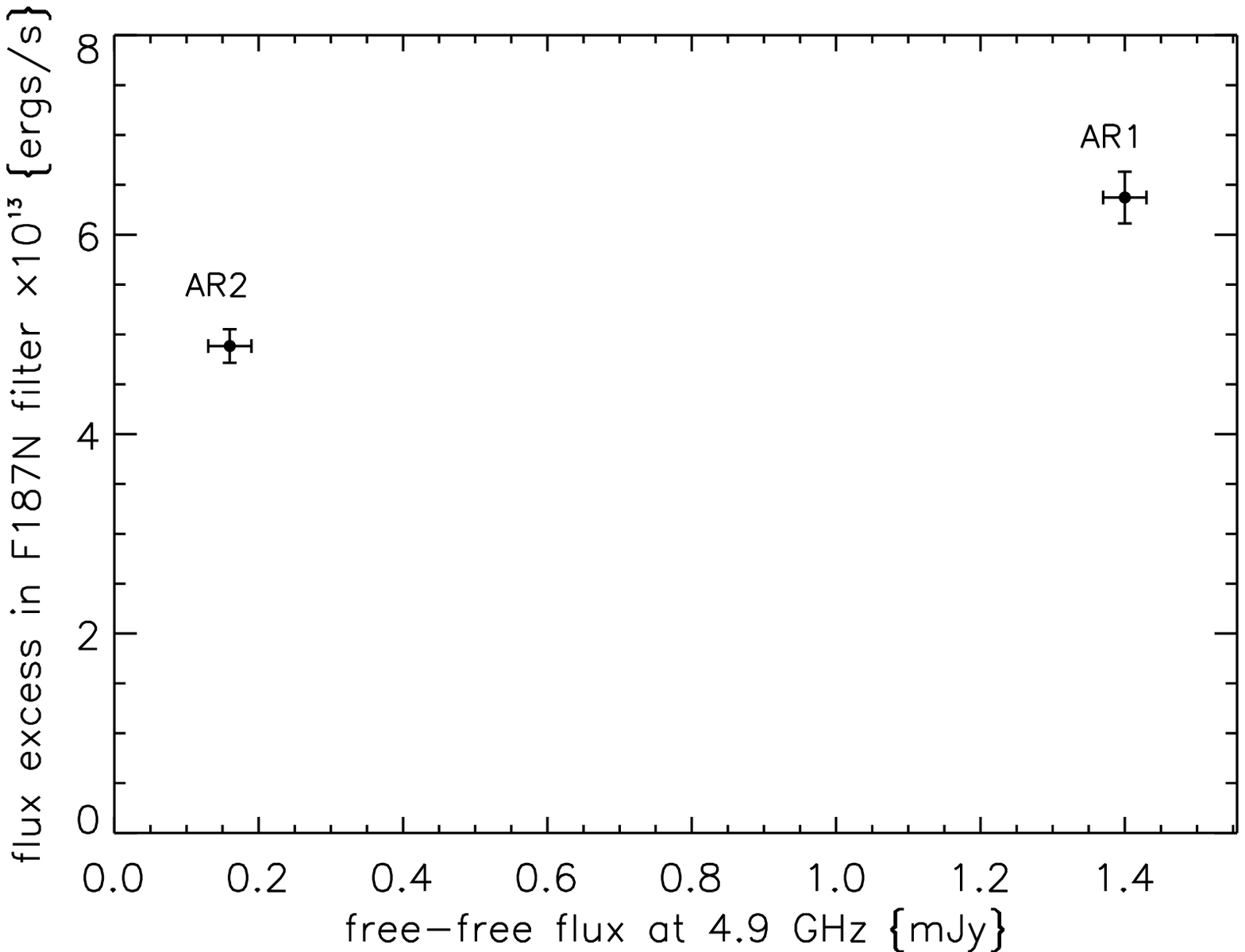}
\caption{Plot of (\Fnp$-\Fnc)\times\Delta\lambda$ versus \Frl.}
\end{figure}


\clearpage 

\begin{figure}
\figurenum{9}
\epsscale{1.05}
\hspace{3.75in}
\plotone{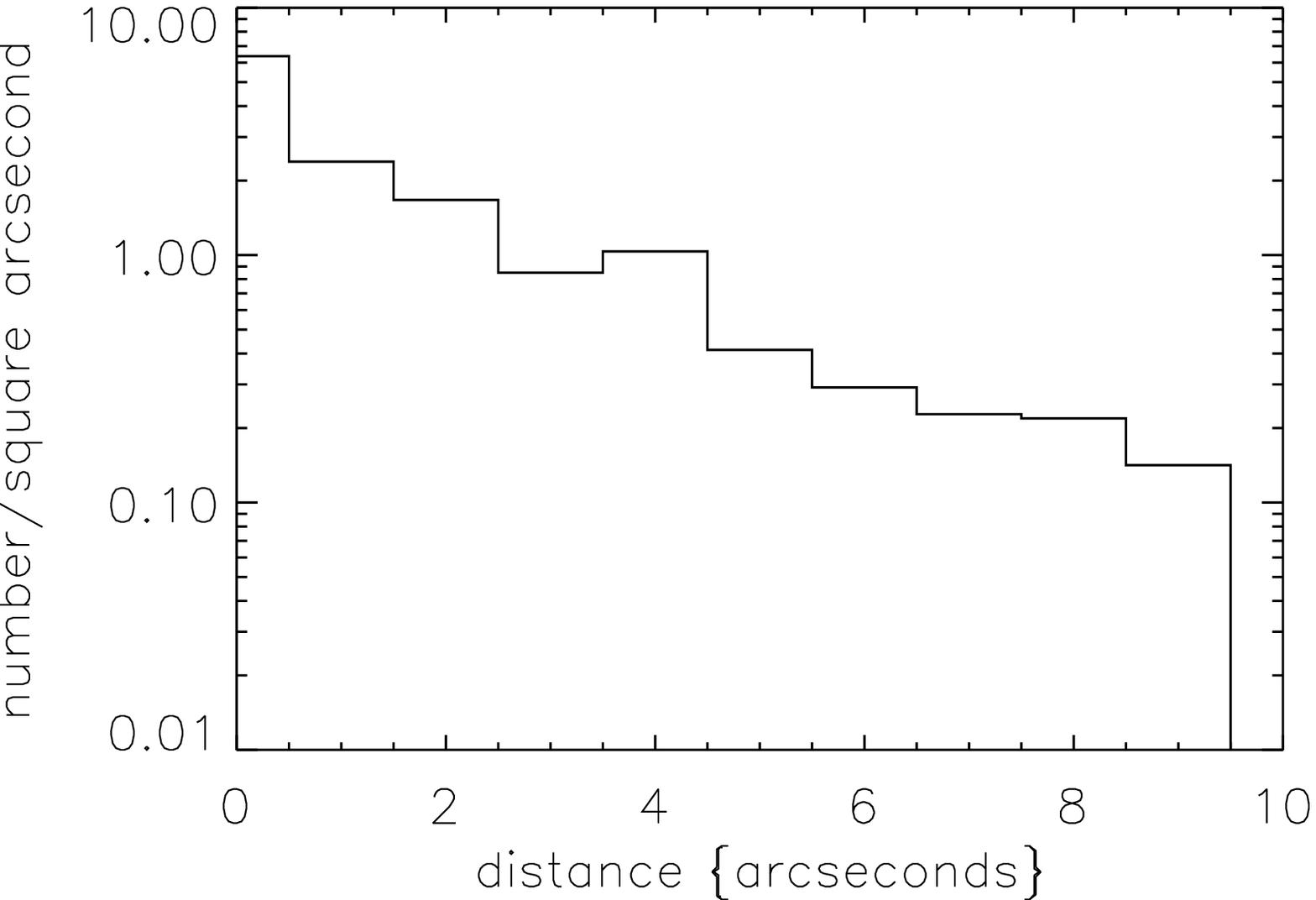}
\hspace*{4.5in} 
\vskip .2in
\caption{Histogram of massive star separations from cluster center. Data are taken from Table 3.}
\end{figure}


\small
\clearpage
\begin{thebibliography}{}
\bibitem[Bieging, Abbott, \& Churchwell(1982)]{bie82} Bieging, J.~H., Abbott, D.~C., \& Churchwell, E.~B.\ 1982, \apj, 263, 207
\bibitem[Blum et al.(2001)]{blu01} Blum, R.\ D., Schaerer, D., Pasquali, A., Heydari-Malayeri, M., Conti, P.\ S., Schmutz, W. 2001, \aj, accepted
\bibitem[Bohannan \& Crowther(1999)]{boh99} Bohannan, B., \& Crowther, P.A., 1999, \apj, 511, 374
\bibitem[Brown \& Liszt(1984)]{bro84} Brown, R.\ L.\ \& Liszt, H.\ S.\ 1984, \araa, 22, 223
\bibitem[Cant{\' o}, Raga, \& Rodr{\' i}guez(2000)]{can00} Cant{\' o}, J., Raga, A.\ C., \& Rodr{\' i}guez, L.\ F.\ 2000, \apj, 536, 896 
\bibitem[Colgan et al.(1996)]{col96} Colgan, S.\ W.\ J., Erickson, E.\ F., Simpson, J.\ P., Haas, M.\ R., \& Morris, M.\ 1996, \apj, 
470, 882 
\bibitem[Conti et al.(1995)]{cont1995} Conti, P. S., Hanson, M. M., Morris, P. W., Willis, A. J., \& Fossey, S. J. 1995, \apj, 445, L35
\bibitem[Cotera et al.(1992)]{cot92} Cotera, A.~S., Erickson, E.~F., Simpson, J.~P., Colgan, S.~W.~J., Allen, D.~A., \& Burton, M.~G.\ 1992, American Astronomical Society Meeting, 181, 8702
\bibitem[Cotera(1995)]{cot95} Cotera, A.\ S.\ 1995, Ph.D.\ Thesis, Stanford University
\bibitem[Cotera et al.(1996)]{cot96} Cotera, A. S., Erickson, E. F., Colgan, S. W. J., Simpson, J. P., Allen, D. A., \& Burton, M. G. 1996, \apj, 461, 750
\bibitem[Crowther \& Dessart(1998)]{cro98} Crowther, P.~A.~\& Dessart, L.\ 1998, \mnras, 296, 622
\bibitem[Davidson et al.(1994)]{dav94} Davidson, J.\ A., Morris, M., Harvey, P.\ M., Lester, D.\ F., Smith, B., \& Werner, M.\ W.\ 
1994, NATO ASIC Proc.\ 445: The Nuclei of Normal Galaxies: Lessons from the Galactic Center, 231 
\bibitem[Eckart et al.(1999)]{eck99} Eckart, A., Ott, T., \& Genzel, R. 1999, A\&A, submitted
\bibitem[Erickson et al.(1991)]{eri91} Erickson, E.~F., Colgan, S.~W.~J., Simpson, J.~P., Rubin, R.~H., Morris, M., \& Haas, M.~R.\ 1991, \apjl, 370, L69
\bibitem[Figer et al.(1998)]{fig98} Figer, D. F., Najarro, F., Morris, M., McLean, I. S., Geballe, T. R., Ghez, A. M., 
\& Langer, N. 1998, \apj, 506, 384
\bibitem[Figer et al.(2000a)]{fig00a} Figer, D.\ F.\ et al.\ 2000a, \apjl, 533, L49
\bibitem[Figer et al.(2000b)]{fig00b} Figer, D.\ F., McLean, 
I.\ S., Becklin, E.\ E., Graham, J.\ R., Larkin, J.\ E., Levenson, N.\ A., \& Teplitz, H.\ I.\ 2000b, \procspie, 4005, 104 
\bibitem[Figer, McLean, \& Najarro(1997)]{fig97} Figer, D.~F., McLean, I.~S., \& Najarro, F.\ 1997, \apj, 486, 420 
\bibitem[Figer, McLean, \& Morris(1995)]{fig95b} Figer, D.~F., McLean, I.~S., \& Morris, M.\ 1995, \apjl, 447, L29
\bibitem[Figer(1995)]{fig95a} Figer, D.\ F.\ 1995, Ph.D.\ Thesis, University of California, Los Angeles
\bibitem[Figer et al.(1999a)]{fig99a} Figer, D.\ F., Kim, S.\ S., Morris, M., Serabyn, E., Rich, R.\ M., \& McLean, I.\ S.\ 1999a, \apj, 525, 750
\bibitem[Figer et al.(1999b)]{fig99b} Figer, D.\ F., Morris, M., Geballe, T.\ R., Rich, R.\ M., Serabyn, E., McLean, I.\ S., Puetter, R.\ C., \& Yahil, A.\ 1999b, \apj, 525, 759 
\bibitem[Figer et al.(2002)]{fig02} Figer, D.\ F., et al.\ 2002, in preparation
\bibitem[Genzel et al.(1990)]{gen90} Genzel, R.\ et al.\ 1990, \apj, 356, 160
\bibitem[Gerhard(2001)]{ger01} Gerhard, O.\ 2001, \apjl, 546, L39
\bibitem[Hanson et al.(1996)]{han96} Hanson, M. M., Conti, P. S., \& Rieke, M. J. 1996, \apjs, 107, 281
\bibitem[Herald, Hillier, \& Schulte-Ladbeck(2001)]{her01} Herald, J.E., Hillier, D.~J.~\& Schulte-Ladbeck, R.E.\ 2001, \apj, 548, 932
\bibitem[Hillier(1989)]{hil89} Hillier, D.~J.\ 1989, \apj, 347, 392
\bibitem[Hillier \& Miller(1998)]{hil98} Hillier, D.~J.~\& Miller, D.~L.\ 1998, \apj, 496, 407 
\bibitem[Hillier \& Miller(1999)]{hil99} Hillier, D.~J.~\& Miller, D.~L.\ 1999, \apj, 519, 354
\bibitem[Ho and Filippenko(1996)]{hof96} Ho, L.~C.~\& Filippenko, A.~V.\ 1996, \apjl, 466, L83 
\bibitem[Illingworth(1976)]{ill76} Illingworth, G.\ 1976, \apj, 204, 73 
\bibitem[Keyes et al.(1997)]{key97} Keyes, T., et al.\ 1997, HST Data Handbook, Ver. 3.0, Vol. I (Baltimore: STScI)
\bibitem[Krabbe et al.(1995)]{kra95} Krabbe, A., et al., \apj, 447, L95
\bibitem[Kim, Morris, \& Lee(1999)]{kim99} Kim, S.\ S., Morris, M., \& Lee, H.\ M.\ 1999, \apj, 525, 228
\bibitem[Kim et al.(2000)]{kim00} Kim, S.\ S., Figer, D.\ F., Lee, H.\ M., \& Morris, M.\ 2000, \apj, 545, 301 
\bibitem[Kim \& Morris(2002)]{kim02} Kim, S.\ S., \& Morris, M.\ 2002, in preparation
\bibitem[Kurucz(1979)]{kur79} Kurucz, R.\ L.\ 1979, \apjs, 40, 1
\bibitem[Lang, Goss, \& Wood(1997)]{lan97} Lang, C.\ C., Goss, W.\ M., \& Wood, D.\ O.\ S.\ 1997, \apj, 481, 1016 
\bibitem[Lang, Goss, \& Rodr{\' i}guez(2001a)]{lan01a} Lang, C.\ C., Goss, W.\ M., \& Rodr{\' i}guez, L.\ F.\ 2001a, \apjl, 551, L143 
\bibitem[Lang, Goss, \& Morris(2001b)]{lan01b} Lang, C.\ C., Goss, W.\ M., \& Morris, M.\ 2001b, \aj, 121, 2681 
\bibitem[Leitherer, Chapman, \& Koribalski(1997)]{lei97} Leitherer, C., Chapman, J.~M., \& Koribalski, B.\ 1997, \apj, 481, 898
\bibitem[McGinn, Sellgren, Becklin, \& Hall(1989)]{mcg89} McGinn, M.~T., Sellgren, K., Becklin, E.~E., \& Hall, D.~N.~B.\ 1989, \apj, 338, 824 
\bibitem[McLean et al.(1998)]{mcl98} McLean et al.\ 1998, SPIE Vol. 3354, 566
\bibitem[McLean et al.(2002)]{mcl02} McLean et al.\ 2002, \pasp, in preparation
\bibitem[Meynet et al.(1994)]{mey94} Meynet, G., Maeder, A., Schaller, G., Schaerer, D., \& Charbonnel, C. 1994, \aap\ Supp., 103, 97
\bibitem[Meynet(1995)]{mey95} Meynet, G.\ 1995, \aap, 298, 767 
\bibitem[Mizutani et al.(1994)]{miz94} Mizutani, K.\ et al.\ 1994, \apjs, 91, 613
\bibitem[Moneti et al.(1994)]{mon94} Moneti, A., Glass, I. S. \& Moorwood, A. F. M. 1994, \mnras, 268, 194
\bibitem[Morris \& Yusef-Zadeh(1989)]{mor89} Morris, M.\ \& Yusef-Zadeh, F.\ 1989, \apj, 343, 703 
\bibitem[Morris, Davidson, \& Werner(1995)]{mor95} Morris, M., Davidson, J.~A., \& Werner, M.~W.\ 1995, ASP Conf.~Ser.~73: From Gas to Stars to Dust, 477
\bibitem[Najarro et al.(2002)]{naj02} Najarro, F.\ et al.\ 2002, in preparation
\bibitem[Nagata et al.(1995)]{nag95} Nagata, T., Woodward, C.\ E., Shure, M., \& Kobayashi, N.\ 1995, \aj, 109, 1676 
\bibitem[Nugis, Crowther, \& Willis(1998)]{nug98} Nugis, T., Crowther, P.~A., \& Willis, A.~J.\ 1998, \aap, 333, 956
\bibitem[Ozernoy, Genzel, \& Usov(1997)]{oze97} Ozernoy, L.~M., Genzel, R., \& Usov, V.~V.\ 1997, \mnras, 288, 237
\bibitem[Perryman et al.(1997)]{per97} Perryman, M.~A.~C.~et al.\ 1997, \aap, 323, L49 
\bibitem[Portegies-Zwart et al.(2001)]{zwa01a} Portegies-Zwart, S.\ F., Makino, J., McMillan, S.\ L.\ W., \& Hut, P.\ 2001, \apjl, 546, L101 
\bibitem[Reid(1993)]{rei93} Reid, M. J. 1993, \araa, 31, 345
\bibitem[Rieke et al.(1989)]{rie89} Rieke, G. H., Rieke, M. J., \& Paul, A. E. 1989, \apj, 336, 752
\bibitem[Salpeter(1955)]{sal55} Salpeter, E. E. 1955, \apj, 121, 161
\bibitem[Scalo(1998)]{sca98} Scalo, J. 1998, in {\it The Stellar Initial Mass Function}, G. Gilmore and D. Howell (eds.), vol. 142 of 38$^{th}$ {\it Herstmonceux Conference}, San Francisco, ASP Conference Series, p. 201
\bibitem[Schaller, Schaerer, Meynet, \& Maeder(1992)]{sch92} Schaller, G., Schaerer, D., Meynet, G., \& Maeder, A.\ 1992, \aaps, 96, 269
\bibitem[Serabyn \& Guesten(1987)]{ser87} Serabyn, E.\ \& Guesten, R.\ 1987, \aap, 184, 133
\bibitem[Serabyn et al.(1998)]{ser98} Serabyn, E., Shupe, D., \& Figer, D. F. 1998, \nat, 394, 448
\bibitem[Serabyn, Shupe, \& Figer(1999)]{ser99} Serabyn, E., Shupe, D., \& Figer, D.~F.\ 1999, ASP Conf.~Ser.~186: The Central Parsecs of the Galaxy, 320 
\bibitem[Serabyn et al.(2002)]{ser02} Serabyn, E., Figer, D.\ F., Kim, S.\ S., Morris, M., \& Rich, R.\ M.\ 2002, \apj, in preparation 
\bibitem[Stetson et al.(1987)]{ste87} Stetson, P. 1987, PASP, 99, 191
\bibitem[Stolte et al.(2002)]{sto02} Stolte, A., Grebel, E.\ K., Brandner, W., \& Figer, D.\ F. 2002, \aa, submitted
\bibitem[Timmermann et al.(1996)]{tim96} Timmermann, R., Genzel, R., Poglitsch, A., Lutz, D., Madden, S.\ C., Nikola, T., Geis, N., \& Townes, C.\ H.\ 1996, \apj, 466, 242 
\bibitem[van der Hucht(2001)]{van01} van der Hucht, K.~A.\ 2001, New Astronomy Review, 45, 135 
\bibitem[Wendker(1995)]{wen95} Wendker, H.~J.\ 1995, \aaps, 109, 177 
\bibitem[Yusef-Zadeh et al.(2001)]{zad01} Yusef-Zadeh, F., Cotera, A., Fruscione, A., Lang, C., Law, C., Wang, D., \& Wardle, M.\ 
2001, American Astronomical Society Meeting, 198, 8709
\bibitem[Yusef-Zadeh et al.(2002)]{zad02} Yusef-Zadeh, F., Law, C., Wardle, M., Wang, Q.\ D., Fruscione, A., Lang, C.\ C., \& Cotera, A.\ 2002, in preparation
\end{thebibliography}
\end{document}